\documentclass[prd,showpacs,amsmath,showkeys,floatfix,amssymb, preprintnumbers, nofootinbib, superscriptaddress]{revtex4}
\usepackage[dvips]{graphics,color}
\usepackage{graphicx}
\usepackage{dcolumn}
\usepackage{bm}
 \usepackage{epsfig}

\usepackage{float}

\newcommand{\eq}{\begin{equation}}
\newcommand{\eeq}{\end{equation}}
\newcommand{\beqa}{\begin{eqnarray}}
\newcommand{\eeqa}{\end{eqnarray}}
\newcommand{\bd}[1]{ \mbox{\boldmath $#1$}  }

\def\ii{\'i}
\begin{document}

\title{Low energy meson spectrum from a QCD approach based on many-body methods}

\author{D. A. Amor-Quiroz}
\email{arturo.amor@nucleares.unam.mx}
\affiliation{
Instituto de Ciencias Nucleares, Universidad Nacional Aut\'onoma de M\'exico
Circuito Exterior, C.U., Apartado postal 70-543, 04510 M\'exico D.F.,
M\'exico.}
\author{T. Y\'epez-Mart\ii nez}
\email{yepez@fisica.unlp.edu.ar} 
\affiliation{Departamento de F\ii sica, Universidad Nacional de La Plata,
C.C. 67 (1900), La Plata, Argentina.}
\author{ P. O. Hess} 
\email{hess@nucleares.unam.mx}
\affiliation{
Instituto de Ciencias Nucleares, Universidad Nacional Aut\'onoma de M\'exico
Circuito Exterior, C.U., Apartado postal 70-543, 04510 M\'exico D.F.,
M\'exico.}

\author{O. Civitarese}
\email{osvaldo.civitarese@fisica.unlp.edu.ar}
\affiliation{Departamento de F\ii sica, Universidad Nacional de La Plata,
C.C. 67 (1900), La Plata, Argentina.}

\author{A. Weber}
\email{axel@ifm.umich.mx}
\affiliation{ Instituto de F\ii sica y Matem\'aticas,
Universidad Michoacana de San Nicol\'as de Hidalgo,
Edificio C-3, Ciudad Universitaria, Apartado Postal 2-82,
58040 Morelia, Michoac\'an, M\'exico}
\date{\today}

\begin{abstract}
The TDA and RPA many-body methods are applied to a
QCD motivated Hamiltonian in the Coulomb gauge. 
The gluon effects in the low energy domain are accounted for 
by the Instantaneous color-Coulomb
Interaction between color-charge densities, approximated by the sum of 
a Coulomb ($\alpha/r$)
and a confining linear ($\beta r$) potentials. 
We use the eigenfunctions of the harmonic oscillator as a basis for the
quantization of the quark fields, and discuss how suitable this basis is 
in various steps of the calculation.
We show that the TDA results already reproduce the gross-structure
of the  light flavored meson states.
The pion-like state in the RPA description, which is a highly
collective state, is in a better agreement with the experimental value.
The results are related to other non-perturbative treatments and compared to
experimental data. We discuss the advantages of the present approach.  
\end{abstract}


\maketitle


\section{Introduction}
\label{sec1}
{\it Quantum Chromodynamics} (QCD) is the theory of the strong interactions. It has
been verified in numerous experiments and is well understood at high energies where, due
to asymptotic freedom, the coupling constant is small and perturbative techniques work well.
At low energy, however, QCD is a highly non-perturbative theory.

Up to now, only {\it Lattice Gauge Theory} (LGT) has been able to obtain reliable results
from first principles (see, {\it e.g.}, \cite{Bietenholtz} for a recent overview).
There are also disadvantages to LGT,  principally the huge numerical effort
involved, but also the difficulties to simulate light quarks with realistic current masses,
to identify the states for a given spin (due to the breaking of rotational symmetry in the
formulation of LGT),  to obtain excited states, etc. 
In recent years, some improvements in the implementation of
LGT have been achieved in describing low lying hadron states 
\cite{Fodor,Aoki} and highly excited meson states \cite{Dudek2011b}.
The Dyson-Schwinger equation (DS) method applied to low energy QCD has 
several advantages, particularly describing in a self-consistent way the
dynamical generation of the constituent quark masses and implementing the
axial-vector Ward-Takahashi identity exactly \cite{Bashir}.
However, the DS method has its own drawbacks, mainly the absence of
clear prescriptions to truncate diagrammatic contributions.

The implementation of schematic models in QCD-motivated Hamiltonians
has provided the complete spectrum of glueball
states \cite{Hess1999}  and hadrons \cite{QCD-I,QCD-III}. 
In all these attempts, it was shown that the use of methods based on
group theory and diagonalization in small spaces can yield
analytic results in certain limits. 
The numerical efforts are less demanding compared to LGT, the rotational
symmetry is maintained, and the identification of the complete
spectrum is feasible.

The formulation of QCD in the Coulomb gauge
\cite{ChristLee,Lee-book} has been widely used. Several systems
described by the non-perturbative regime of QCD have been examined in
this framework. The glueball spectrum, studied by LGT \cite{Bali,Morningstar}, was well
reproduced in \cite{Adam1996,Adam2003}. The calculations for systems involving heavy quarks
\cite{Peng}, like charmonium and bottomonium, have approximately reproduced the available data.
The charmonium hybrid meson spectrum and their radiative transitions, including controversial
exotic states, were calculated  in \cite{Peng,Yepez2014} and compared
to LGT results \cite{Dudek2006,Dudek2009}.
Finite temperature studies focused on the Yang-Mills sector of the theory
\cite{Hugo2011,Yepez2012} have been able to describe the possibility of a phase
transition for this system, as well as its thermal properties described before by LGT \cite{Engels,Boyd}.

As a starting point for a different approach to diagonalizing the QCD Hamiltonian
that employs the effective formulation developed in \cite{Adam2001},
analytic results were obtained in \cite{Hess2006} for light quark
systems restricted to one orbital level. This approach was extended to
an arbitrary number of orbital levels by applying a diagonalization to
the kinetic energy of the pure quark sector of the QCD Hamiltonian
in a finite volume \cite{Yepez2010}. 
It was shown that the use of the harmonic oscillator basis has several practical
advantages as it leads to analytical expressions for the relevant matrix elements.
In this basis it is also straightforward to subtract the contributions of
the center of mass motion. Though the harmonic oscillator states
represent a non-relativistic basis, which implies that one has to use
many states to approximate relativistic states, its
use is advantageous. Recently, we have investigated the
implementation of many-body methods in a $SO(4)$-model 
for quarks \cite{So4-I,So4-II}, which leads to a pion-like state with high collectivity.

In this contribution we go further and try to extract information
about the microscopic structure of low energy meson states, by
approximating the QCD Instantaneous color-Coulomb Interaction (QCD-IcCI) which has been derived in
\cite{ChristLee,Lee-book}.  
In \cite{Adam2001,Hugo2004}, the momentum dependence of the QCD-IcCI 
has been studied for a static quark-antiquark pair, 
using a self-consistent mean field approximation.
It was shown that   
in coordinate space it corresponds to a linear potential, of the type found in LGT. 
Motivated by these results, we substitute the QCD-IcCI by a Coulomb plus linear potential, 
the latter being the confining part of the quark-quark interaction.
This potential acts instantaneously and its use may be questioned 
for light quarks. In any case, the effective potential can only be considered
as a first approximation, and the effects of dynamical gluons should
eventually be included \cite{Adam2001}.
The main objective of the present work is to show that there is a possibility
to improve and significantly simplify the burden of calculations in
low-energy non-perturbative QCD by using many-body methods, and that the harmonic
oscillator basis is well suited for this task. 
The strategy will be explained in detail in the course of the paper.

The paper is organized as follows: in Section \ref{sec2}, the effective QCD
Hamiltonian derived in \cite{Adam2001} will be discussed and the
approximations will be justified.  The Hamiltonian is then rewritten in terms of
the harmonic oscillator basis and its operators. 
A prediagonalization is applied to the kinetic plus mass terms,
after which it is possible to define effective quark and antiquark
operators. In Section \ref{TDA-RPA}, the Tamm-Dancoff-Approximation  (TDA) and the
Random-Phase-Approximation (RPA) are
applied to the Hamiltonian, and the spectrum of meson states
viewed as collective states is calculated. 
We will show that the meson spectrum is, to a large extent, reproduced.  
While the energy of the pion-like state in TDA is higher than the experimental data,  
the RPA result for the same state is in better agreement with the experimental value. 
We also compare our results to the spectrum obtained in \cite{Llanes}, 
where a continuum basis for the description of the meson states was used
and the same approximations were applied.
Since the RPA method introduces pair correlations in the ground state,
the interactions that account for them are identified and their
effects quantified. In subsection \ref{Sec:RN}, the  renormalization of the
quark masses and couplings is described.
We present our conclusions in Section \ref{Conclusions}. The technical
details of the approach are relegated to five appendices.

\section{The QCD Hamiltonian at low energy}
\label{sec2}
As we have already mentioned in Section \ref{sec1}, QCD has been widely studied
in its canonical Coulomb gauge representation. In Ref. \cite{Adam2001}, it was shown how
confinement and the constituent particles (quarks and gluons) can be treated simultaneously 
within this gauge. The QCD Hamiltonian in the canonical Coulomb gauge representation is
given by \cite{ChristLee,Lee-book}
\beqa
{\bd H}^{QCD}&=&\int \left\{ \frac{1}{2} 
\left[ \mathcal{J}^{-1}  \mathcal{ {\bf \Pi}}^{tr }  \cdot  \mathcal{J} \mathcal{ {\bf \Pi}}^{tr }
+ {\bf \mathcal{B} }  \cdot  {\bf \mathcal{  B}}\right]
-\overline{{ \psi}}\left(-i{ \gamma}\cdot{ \nabla}+m\right) { \psi}
-g\overline{{ \psi}}{ \gamma}\cdot { A}
{ \psi}\right\}d {\bf  x}
\nonumber \\
&& +\frac{1}{2}g^{2}\int\mathcal{J}^{-1}\rho^{c}( {\bf  x})
\langle{c,{\bf  x}}|\frac{1}{{ \nabla}\cdot \mathcal{{ D}}}
(-{ \nabla}^{2})\frac{1}{{ \nabla}\cdot \mathcal{{ D}}}
|c^{\prime} {\bf y}\rangle
\mathcal{J}\rho^{c^{\prime}}({\bf  y})d {\bf x}d {\bf y}~.
\label{eq1}
\eeqa

Here, $\mathcal{ {\bf \Pi}}^{tr }$ and $ {\bf \mathcal{B} }$ 
are the transverse chromo-electromagnetic 
fields in the QCD Coulomb gauge and  ${ \psi}$ represents the quark fields. 
The last two terms in Eq.\ (\ref{eq1}) are the quark-gluon
interaction ($g$-term) and the total 
color-charge density  interaction ($g^2$-term),
respectively. In QCD, the total color-charge density $\rho^c({\bf x})$ 
contains the contribution of quarks-antiquarks and gluons.
The kernel in the last term of Eq.\ (\ref{eq1}) is the 
Instantaneous color-Coulomb Interaction in QCD (QCD-IcCI)
which includes the inverse of the Faddeev-Popov operator 
$({ \nabla}\cdot \mathcal{{ D}})^{-1}$ 
and its determinant $\mathcal{J}=\mbox{det}({ \nabla}\cdot
\mathcal{{  D}})$ \cite{ChristLee,Lee-book}. 
At low energy, light quarks play the most important role, while the
effects of dynamical gluons can be simulated by the interaction 
$V(| {\bf x} -{\bf y} |)=-\frac{\alpha}{| {\bf x} -{\bf y}|}+\beta |{\bf x} -{\bf y}| $,
which is obtained from a self-consistent treatment of the
interaction between color charge-densities \cite{Adam2001,Hugo2004}.

Therefore, an approximate effective QCD Hamiltonian can be written as 
\beqa\label{Eff-QCD-H}
{\bd H}^{QCD}_{eff} & = &
\int\left\{
{\bd \psi}^{\dagger}({\bf x})
(-i{\bd \alpha}\cdot{\bd \nabla}+\beta{m})
{\bd \psi}({\bf x})\right\}d{\bf x}
-\frac{1}{2}
\int\rho_{c}({\bf x})
V(|{\bf x}-{\bf y}|)\rho^{c}({\bf y})d{\bf x}d{\bf y}
\nonumber \\
& = & {\bd K} + {\bd H}_{{\rm Coul}}~,
\label{eq2}
\eeqa
where we have restricted ourselves to the quark sector of the theory, without
dynamical gluons. Consequently, 
$\rho^c({\bf x}) = \psi^\dagger({\bf x}) T^c \psi({\bf x})$
now represents the quark and antiquark charge density.
In the last line of Eq.\ (\ref{Eff-QCD-H}), the first term is the kinetic energy,
while the second term is the QCD-IcCI in its simplified form.
The reason for the factor of $-\frac{1}{2}$ in the potential
term is that the short-range interaction of two quarks with
opposite (color) charges is similar to the classical attractive Coulomb interaction, i.e., $-\frac{q^2}{r}$
\cite{Adam1996,Llanes2000}. 
The fields $\psi^\dagger ({\bf x})$ and its hermitian conjugate 
are quantized in the usual way, by expanding them in terms of  creation and annihilation
operators and using as coefficients the harmonic oscillator functions:
\beqa\label{fermion-field}
\psi^\dagger ({\bf x}) & = & \sum_{\tau,N l m_l,\sigma c f} 
R^{*}_{Nl}(x) Y^*_{lm_l}(\hat {\bf x}) \chi^\dag_{\sigma} ~{\bd q}^\dag_{\tau, Nlm_l, \sigma c  f}\nonumber\\
& = & \sum_{Nlm_l, \sigma c f}   R^{*}_{Nl}(x) Y^*_{lm_l}(\hat {\bf x}) \chi^\dag_{\sigma}
\left( {\bd q}^\dag_{\frac{1}{2}, Nlm_l ,\sigma c f} 
+  {\bd q}^\dag_{-\frac{1}{2}, Nlm_l,\sigma c f}\right) ~. 
\label{quant}
\eeqa
with $x=| {\bf x} |$ and  $ R_{Nl}(x)=\textit{N}_{Nl}\exp(-\frac{B_0 x^{2}}{2})
x^{l}L_{\frac{N-l}{2}}^{l+\frac{1}{2}}(B_0 x^{2})$, 
where $L^\lambda_n$ is an associated  Laguerre polynomial
and $( \sqrt{B_0})^{-1}$ is the oscillator length. 
The index $\tau$ denotes the upper ($\tau=\frac{1}{2}$) and lower ($\tau=-\frac{1}{2}$) 
components of the Dirac spinors  in the Dirac-Pauli representation of the Dirac matrices,
and $\sigma,c,f$ indicate the spin, color and flavor intrinsic degrees of freedom, respectively.

For the implementation of the {\it Tamm-Dancoff Approximation} (TDA) and the {\it Random Phase Approximation} 
(RPA) methods in the present approach, we have developed the following strategy:
in a first step, the kinetic energy term is diagonalized within a finite volume (Section \ref{Sec:Kinetic-term}), 
which we shall refer to as ``prediagonalization''.  The eigenfunctions constitute 
a new basis, where  quark and antiquark creation and annihilation operators can be defined. 
In a second step (Section \ref{Sec:eff-int}), the interaction is rewritten in terms of these
creation and annihilation operators. 
Finally, the whole Hamiltonian of Eq.\ (\ref{Eff-QCD-H}), expressed in the effective basis, is
diagonalized in the space of particle-hole collective states (Section \ref{TDA-RPA}).

\subsection{Diagonalization of the kinetic energy and the effective basis.}
\label{Sec:Kinetic-term}

The kinetic energy in terms of the creation and annihilation operators of the
oscillator basis,  in their total spin ${\bf j}={\bf l}+ \frac{1}{2}$ representation,
is written as
\beqa
{\bd K} & = & \sum_{j\tau_i N_i l_i} 
\sum_{m c f}  K^{j,T}_{\tau_1(N_1l_1),\tau_2(N_2l_2)}
{\bd q}^\dagger_{\tau_1(N_1 l_1)jmcf}  {\bd q}^{\tau_2(N_2 l_2)jmcf}
~~~,
\label{eqq3}
\eeqa
which is clearly not diagonal. 
The indices  $m$, $c$ and $f$ correspond to the 
total spin $m=-j,\cdots ,j$, color $c=1,2,3$ and flavor $f=u,d,s$
components, respectively, where $f=\{Y,T,T_z\}$ is a short-hand notation
for flavor-hypercharge, isospin and the third component of isospin.
We distinguish between the u-d quarks ($T=\frac{1}{2}$) and the s
quarks ($T=0$), where the latter have a larger mass.
The prediagonalization has to be applied in each sector with a
different mass or isospin {\it i.e.}, 
$m^T_0= m_{u,d} \delta_{T,  \frac{1}{2}}+ m_s \delta_{T,0} $. 
The matrix $K^{j,T}_{\tau_1(N_1l_1),\tau_2(N_2l_2)}$ (given explicitly
in Appendix \ref{appA}) has to be diagonalized in order to obtain 
an effective basis with respect to which the total Hamiltonian is then diagonalized.

The diagonalization of the kinetic term (\ref{eqq3}) is performed for a given 
maximal number of quanta $N=N_{{\rm cut}}$, for which we introduce a general transformation
to a basis of effective operators,
\beqa
{\bd q}^\dagger_{\tau (N l)jmcf} & = &
\sum_{\lambda \pi k} \left( \alpha^{j,T}_{\tau (Nl),\lambda \pi k}\right)^* 
{\bd Q}^\dagger_{\lambda \pi kjmcf}  
~\delta_{\pi,(-1)^{\frac{1}{2}-\tau+l}}
~~~.
\label{eq3}
\eeqa
The index $\lambda = \pm\frac{1}{2}$ refers 
to the pseudo-spin components 
after the diagonalization of the kinetic term, and
$k$ runs over all orbital states after  the  diagonalization. 
The value $\lambda = +\frac{1}{2}$ refers to positive energy states
(effective quarks) and the value
$\lambda = -\frac{1}{2}$ to negative energy states (effective antiquarks). 
For example, taking $N_{{\rm cut}}=3$ and
$j=\frac{1}{2}$, only the lowest two $s$ and $p$ orbital states contribute,
so that the index for the positive energy states runs from $k$ = 1 to 2, for a fixed $\tau$.
The same happens for  the negative energy states.
The parity on the left-hand side (L.H.S) in (\ref{eq3}) is given by
$ (-1)^{\frac{1}{2}-\tau+l}$  while for the right-hand side (R.H.S.)
$\pi$ indicates the parity after the diagonalization {\it i.e.}, our
effective particles $({\bd Q}^\dagger_{\lambda \pi kjmcf})$ are a
linear combination of states with the same well-defined parity. 
The transformation coefficients depend on the type of quarks, whether it is
an up or down quark (equal masses are assumed) or a strange quark ($m_{u,d}<m_s$). 
For the transformation coefficients and the kinetic matrix elements,
only the dependence on the flavor isospin is given,
because the flavor-hypercharge is fixed by $T$.

The eigenvalue problem to be solved acquires the following form
\beqa\label{eq:prediag}
\sum_{\tau_i  N_i l_i } 
 \alpha^{j,T}_{\tau_1,(N_1 l_1),   \lambda_1 \pi_1 k_1} 
K^{j,T}_{\tau_1(N_1l_1),\tau_2(N_2l_2)}
 \alpha^{j,T}_{\tau_2,(N_2 l_2), \lambda_2 \pi_2 k_2} 
=\varepsilon_{\lambda_1 \pi_1 k_1 j Y T}
~\delta_{\lambda_1 \lambda_2}\delta_{\pi_1 \pi_2}\delta_{k_1 k_2}~,
\eeqa
where we have  taken the transformation coefficients of Eq.\ (\ref{eq3}) to be real.

It is worth mentioning that the spin, color and flavor quantum numbers
of the single particles are conserved in the prediagonalization. 
In  this work, we only consider single particles with total spin
$j=\frac{1}{2}$ in order to study low energy states built by pairs coupled to spin $J=0,1$
and leave for future work the analysis of the contributions due to
single particle spin $j=\frac{3}{2}$. The method presented here can directly be extended to $j > \frac{1}{2}$.

Having performed the diagonalization of Eq.\ (\ref{eqq3}), 
the kinetic energy is rewritten in terms of effective
creation (annihilation) operators of 
quarks ${\bd b}^\dagger({\bd b}$) and antiquarks ${\bd d}^\dagger({\bd  d}$) 
(see Appendix \ref{appA}),
\beqa
{\bd K} &   =   & \sum_{\pi k j Y T} \varepsilon_{\pi k j Y T} 
\sum_{mcT_z}\left( {\bd b}^\dagger_{\pi k j Y T, m c T_z} {\bd b}^{\pi k j Y T,m c T_z}
- {\bd d}^{\pi k j Y T, m c T_z}  {\bd d}^\dagger_{\pi k j Y T, m c T_z}
\right)
~~~, \label{eq9}
\eeqa
where the eigenvalues (effective masses) are denoted by
$\varepsilon_{\pi k j Y T}$ and the quark and antiquark operators are related
to the operators before the prediagonalization via 
${\bd Q}^\dagger_{\frac{1}{2} \pi kjmcf}  \rightarrow  {\bd  b}^\dagger_{\pi k j Y T, m c T_z}$
and
${\bd Q}^\dagger_{-\frac{1}{2} \pi kjmcf}  \rightarrow  {\bd d}_{\pi  k j Y T, m c T_z}$ 
(see (\ref{eq4})).

The diagonalization is achieved by 
choosing the oscillator length $(\sqrt{B_0})^{-1}$ in such a way that for a given cut-off $N_{{\rm cut}}=N_0$
(called the {\it renormalization point}, see Appendix E)
the first ($k=1$)  single-particle state $\varepsilon_{\pi k j Y T}$
is at a fixed energy. 
This energy is chosen such that it represents a confined particle in a finite volume of the
size of a hadron, {\it i.e.}, $0.5\mbox{fm} \le (\sqrt{B_0})^{-1} \le 1\mbox{fm}$. 
The renormalization of the masses is explained in the Section \ref{Sec:RN} and in the Appendix \ref{Ap-RN}.

\subsection{The effective Coulomb Interaction}
\label{Sec:eff-int}
In terms of the creation ${\bd q}^\dagger_{\tau (Nl)jmfc}$ and
annihilation ${\bd q}^{\tau (Nl)jmfc}$ operators, before the prediagonalization,
the effective Coulomb interaction acquires the form (see Appendix B)
\beqa
{\bd H}_{{\rm Coul}}
&=& 
-\frac{1}{2}   \sum_{N_i l_i j_i L} V_{\{N_i l_i j_i\}}^{L}
\left[
\left[   {\bd q}^\dagger_{(N_1l_1)j_1}  \otimes  {\bd q}_{(N_2l_2)j_2}  \right]^{0,L,(11),(00)}
\otimes
\left[  {\bd q}^\dagger_{(N_3l_3)j_3}  \otimes  {\bd q}_{(N_4l_4)j_4}  \right]^{0,L,(11),(00)}
\right]^{0,0,(00),(00)}_{0,0,~0,~~0}   ,
\label{eq10}
\eeqa

The intermediate coupling refers to pseudo-spin zero, angular momentum $L$, color (1,1) and flavor
(0,0), respectively.  Since the TDA and RPA 
meson-like pairs will not be coupled to a definite $SU(3)$ 
flavor irrep but only to a definite $Y,T,T_z$, it is preferable to rewrite the product 
$\left[{\bd q}^\dagger_{(N_1l_1)j_1}\otimes {\bd q}_{(N_2l_2)j_2} \right]^{0,L,(11),(00)}_{0, M_L, C, 0}$ 
as
\beqa
\left[ {\bd q}^\dagger_{(N_1l_1)j_1}
\otimes  {\bd q}_{(N_2l_2)j_2}
\right]^{0,L,(11),(00)}_{0,M_L, C, 0}
=  \sum_{Y_1T_1,Y_2T_2}
(-1)^{\frac{1}{3}+\frac{Y_1}{2}+T_1}\frac{\sqrt{2T_1+1}}{\sqrt{3}}\delta_{T_2T_1}\delta_{Y_2Y_1}
\left[  {\bd q}^{\dagger}_{  (N_1 l_1 ) j_1 Y_1 T_1 } 
\otimes  {\bd q}_{  (N_2 l_2 ) j_2 \bar{Y}_2 T_2 }  
\right]^{0,L, (11),00}_{0,M_L ,C ,0}
~,\nonumber\\
\label{eq10-1}
\eeqa
where $\bar Y_2=-Y_2$ and the new factor in the sum comes from an isoscalar factor of
$SU(3)$ \cite{jutta}. The last two zeros in the final coupling indices refer to $Y$ and $T$
while the (last) lower magnetic quantum number index refers to $T_z$.

Applying the transformation (\ref{eq3}),
the product of a creation and an annihilation operator in (\ref{eq10-1}) transforms to
\beqa
&&\left[ 
{\bd q}^{\dagger}_{  (N_1 l_1 ) j_1 Y_1 T_1 } 
\otimes {\bd q}_{  (N_2 l_2 ) j_2 \bar{Y}_2 T_2 }  
\right]^{0,L, (11),00}_{0,M_L, C ,0}\nonumber\\
&&= 
\frac{1}{\sqrt{2}} \sum_{\tau_1, \tau_2}~ \sum_{\lambda_1 \pi_1 k_1,  \lambda_2\pi_2 k_2 } 
\delta_{\tau_1,\tau_2}~ \delta_{ \pi_1 , (-1)^{\frac{1}{2}-\tau_1 + l_1} }~
\delta_{  \pi_2 , (-1)^{\frac{1}{2}-\tau_2 + l_2} }
\left(  \alpha^{j_1,T_1}_{\tau_1 ( N_1 l_1 ), \lambda_1 \pi_1 k_1}  \right)
\left(  \alpha^{j_2, T_2}_{\tau_2 ( N_2 l_2 ), \lambda_2 \pi_2 k_2}  \right)
\nonumber \\
&&\times~~
\bigg\{\delta_{\lambda_1,\frac{1}{2}} \delta_{\lambda_2,\frac{1}{2}}
\left[  {\bd b}^\dagger_{ \pi_1 k_1 j_1 Y_1 T_1 } 
\otimes  {\bd b}_{  \pi_2  k_2 j_2 \bar{Y}_2 T_2 } \right]^{L, (11),00}_{M_L,C, 0}
-
\delta_{\lambda_1,\frac{1}{2}} \delta_{\lambda_2,-\frac{1}{2}}
\left[  {\bd b}^\dagger_{ \pi_1 k_1 j_1 Y_1 T_1 } 
\otimes 
{\bd d}^{\dagger}_{ \pi_2  k_2 j_2 \bar{Y}_2 T_2 } \right]^{L, (11),00}_{M_L,C ,0}
\nonumber\\
&&+~~ \delta_{\lambda_1,-\frac{1}{2}} \delta_{\lambda_2,\frac{1}{2}}
\left[ {\bd d}_{ \pi_1 k_1 j_1 Y_1 T_1 } 
\otimes  {\bd b}_{ \pi_2  k_2 j_2 \bar{Y}_2 T_2 } \right]^{L, (11),00}_{M_L, C ,0}
-
\delta_{\lambda_1,-\frac{1}{2}} \delta_{\lambda_2,-\frac{1}{2}}
\left[ {\bd d}_{ \pi_1 k_1 j_1 Y_1 T_1 } 
\otimes  {\bd d}^\dagger_{ \pi_2  k_2 j_2 \bar{Y}_2 T_2 } \right]^{L, (11),00}_{M_L, C, 0}
\bigg\}
~~~.
\label{eq11}
\eeqa

This allows us to introduce the following notation
\beqa
\mathcal{F}_{1, 2;~\Gamma_0, \mu_0 }
&=&
\frac{1}{\sqrt{2}} \bigg\{ \delta_{\lambda_1,\frac{1}{2}} \delta_{\lambda_2,\frac{1}{2}}
\left[ 
{\bd b}^\dagger_{ \pi_1 k_1 j_1 Y_1 T_1 }  \otimes  {\bd b}_{  \pi_2  k_2 j_2 \bar{Y}_2 T_2 }
\right]^{L, (11),00}_{M_L, C ,0}
-
\delta_{\lambda_1,-\frac{1}{2}} \delta_{\lambda_2,-\frac{1}{2}}
\left[ {\bd d}_{ \pi_1 k_1 j_1 Y_1 T_1 }   \otimes   {\bd d}^\dagger_{ \pi_2  k_2 j_2 \bar{Y}_2 T_2 }
\right]^{L (11)00}_{M_L, C, 0} \bigg\}
\nonumber\\
\mathcal{G}_{1, 2;~\Gamma_0,\mu_0}
&=& \frac{1}{\sqrt{2}}  \bigg\{
\delta_{\lambda_1,-\frac{1}{2}} \delta_{\lambda_2,\frac{1}{2}}
\left[   {\bd d}_{ \pi_1 k_1 j_1 Y_1 T_1 }   \otimes  {\bd b}_{  \pi_2  k_2 j_2 \bar{Y}_2 T_2 }
\right]^{L, (11),00}_{M_L, C, 0}
-
\delta_{\lambda_1,\frac{1}{2}} \delta_{\lambda_2,-\frac{1}{2}}
\left[ {\bd b}^\dagger_{ \pi_1 k_1 j_1 Y_1 T_1 }  \otimes  {\bd d}^\dagger_{  \pi_2  k_2 j_2 \bar{Y}_2 T_2 }
\right]^{L, (11),00}_{M_L,C, 0}     \bigg\}
~~~.\nonumber\\
\label{eq12}
\eeqa
where we have used the short-hand notations
$1=\lambda_1\pi_1 k_1 j_1 Y_1 T_1 $ (similarly for the index $2$),
for the quantum numbers of the  intermediate coupling in the interaction $\Gamma_0= L,(11),00$, and for
their magnetic projections $\mu_0=M_L,C,0$, respectively.

With this, the Coulomb interaction is rewritten as
\beqa
\bd{H}_{Coul} &=& - \frac{1}{2} \sum_{L }
\sum_{ \lambda_i \pi_i  k_i j_i  Y_i T_i  } 
V^{L}_{ \{\lambda_i  \pi_i  k_i  j_i Y_i T_i \} } 
\Big(   \left[\mathcal{F}_{12;\Gamma_0}
\mathcal{F}_{34;\bar \Gamma_0}\right]^{\tilde 0}_{\tilde 0}
+\left[\mathcal{F}_{12;\Gamma_0}\mathcal{G}_{34;\bar \Gamma_0}\right]^{\tilde 0}_{\tilde 0}
+\left[\mathcal{G}_{12;\Gamma_0}\mathcal{F}_{34;\bar \Gamma_0}\right]^{\tilde 0}_{\tilde 0}
+\left[\mathcal{G}_{12;\Gamma_0}\mathcal{G}_{34;\bar \Gamma_0}\right]^{\tilde 0}_{\tilde 0}
\Big)~,\nonumber\\
\label{eq13}
\eeqa
where the upper index $\tilde 0=\{0,(00),00\}$ indicates the total
couplings of the interaction in spin, color and flavor hypercharge and
isospin, while the lower index $\tilde 0=\{0,0,0\}$ indicates the corresponding 
magnetic numbers.

The second and third terms in Eq. (\ref{eq13}) do not contribute in either
the TDA or the RPA scheme, since they have an odd number of creation and
annihilation operators.  They can eventually be absorbed in higher order terms of the coupling
between quarks, antiquarks and phonons. The creation and annihilation of two pairs,
contained in the last term of Eq. (\ref{eq13}), only contributes in the
RPA scheme and accounts for the presence of ground state correlations in this framework.

The matrix elements in Eq.\ (\ref{eq13}) are given by
\beqa\label{eq-new-matrix-elements}
V^{L}_{ \{\lambda_i \pi_i k_i  j_i Y_i T_i \} } 
&=&\sum_{\tau_i N_i l_i}~ V_{\{N_i l_i j_i\}}^{L}~
\alpha^{j_1,T_1}_{\tau_1(N_1l_1),\lambda_1,\pi_1,k_1}
\alpha^{j_2,T_2}_{\tau_2(N_2l_2),\lambda_2,\pi_2,k_2}
\alpha^{j_3,T_3}_{\tau_3(N_3l_3),\lambda_3,\pi_3,k_3}
\alpha^{j_4,T_4}_{\tau_4(N_4l_4),\lambda_4,\pi_4,k_4} \nonumber\\
&\times&
~\delta_{\tau_1\tau_2} \delta_{\tau_3\tau_4}~
\delta_{ \pi_1 , (-1)^{\frac{1}{2}-\tau_1 + l_1} }
\delta_{  \pi_2 , (-1)^{\frac{1}{2}-\tau_2 + l_2} }
\delta_{ \pi_3 , (-1)^{\frac{1}{2}-\tau_3 + l_3} }
\delta_{  \pi_4 , (-1)^{\frac{1}{2}-\tau_4 + l_4} } \nonumber\\
& \times&
(-1)^{\frac{1}{3}+\frac{Y_1}{2}+T_1}\frac{\sqrt{2T_1+1}}{\sqrt{3}}\delta_{T_2T_1}\delta_{Y_2 Y_1}
~(-1)^{\frac{1}{3}+\frac{Y_3}{2}+T_3}\frac{\sqrt{2T_3+1}}{\sqrt{3}}\delta_{T_4T_3}\delta_{Y_4 Y_3}~,
\eeqa
and the matrix elements in the harmonic oscillator basis ($V_{\{N_i l_i j_i\}}^{L}$) are analytic and 
actually easy to compute. They are given in Appendix \ref{appB}.

\section{The TDA and RPA bosonization methods.}
\label{TDA-RPA}
The TDA and RPA are both bosonization (or linearization)
methods \cite{Ring}. In Appendix \ref{AP:TDA-and-RPA}, we present a brief
introduction to the RPA method and its simplified TDA limit. 
The quark-antiquark pair operators needed for the TDA and RPA one
phonon states with quantum numbers $\Gamma=\{J^P,(0,0)_C,Y,T\}$ and magnetic projection
numbers $\mu=\{M_j, 0, T_z\}$ are given by
\beqa
[{\bd b}^\dag_{\pi_a k_a j_a Y_a,T_a} \otimes 
{\bd d}^\dag_{\pi_b k_b j_b \bar  Y_b,T_b} ]^{J^P,(0,0)_C,Y,T}_{M_J,~ 0, ~T_z}
=[{\bd b}^\dag_{\bf a} {\bd d}^\dag_{\bar {\bf b}} ]^\Gamma_\mu ~,
\label{pair-states}
\eeqa
where the creation (annihilation) operators of effective particles 
${\bd b}^\dag_{\pi_i k_i j_i m_i c_i f_i}({\bd b}^{\pi_i k_i j_i m_i  c_i f_i})$ and antiparticles
${\bd d}^{\dag \ \pi_i k_i j_i m_i  c_i f_i}({\bd d}_{\pi_i k_i j_i  m_i c_i f_i})$ 
were introduced in the previous Section.
On the R.H.S. of Eq.\ (\ref{pair-states}) we have introduced the short-hand notation that we are going to use from now
on, in order to describe all possible pairs needed to construct the collective TDA and RPA phonon operators.

Here, we present the more general phonon operators that we are going to implement, {\it i.e.}, the RPA
creation phonon operators $\hat \Gamma^\dagger_{n;\Gamma\mu}$, which are defined as
\beqa
\hat \Gamma^\dag_{n;\Gamma\mu}
&=&\sum_{{\bf a}, {\bf b}} \left\{X^n_{{\bf a}{\bf b};\Gamma}  
[{\bd b}^\dag_{\bf a} {\bd d}^\dag_{\bar {\bf b}} ]^\Gamma_\mu
-Y^n_{{\bf a}{\bf b};\Gamma}  (-1)^{\phi_{\Gamma\mu}}
[{\bd d}^{\bar {\bf b}}  {\bd b}^{ {\bf a}}  ]^{\bar \Gamma}_{\bar \mu}\right\}\nonumber\\
&=&\sum_{{\bf a}, \mu_{\bf a}} \sum_{ {\bf b},\mu_{\bf b}} \left\{X^n_{{\bf a} {\bf b};\Gamma}  
~\langle {\bf a} \mu_{\bf a}, \bar {\bf b} \bar \mu_{\bf b}| \Gamma\mu \rangle 
~{\bd b}^\dag_{\bf a \mu_{\bf a}}  {\bd d}^\dag_{\bar {\bf b} \bar \mu_{\bf b}}  
-Y^n_{{\bf a}{\bf b};\Gamma}  (-1)^{\phi_{\Gamma\mu}}
~\langle  {\bf a} \mu_{\bf a},\bar {\bf b} \bar \mu_{\bf b} |  \bar \Gamma \bar \mu \rangle 
~{\bd d}^{\bar {\bf b} \bar \mu_{\bf b}}  {\bd b}^{ {\bf a} \mu_{\bf a}} 
\right\}  \label{eq15}
\eeqa
where $(-1)^{\phi_{\Gamma\mu}}=(-1)^{J-M_J}(-1)^{T-T_z}$ is the phase
needed to guarantee the scalar  nature of the vacuum.

The corresponding annihilation phonon operators $\hat
\Gamma^{n;\Gamma\mu}$ are 
\beqa
\hat \Gamma^{n;\Gamma\mu}
&=&\sum_{{\bf a},{\bf b}}
\left\{ \left(X^n_{{\bf a}{\bf b};\Gamma}\right)^*
[{\bd d}^{\bar {\bf b}}  {\bd b}^ {\bf a}]^\Gamma_\mu
-\left(Y^n_{{\bf a}{\bf b};\Gamma}\right)^* (-1)^{\phi_{\Gamma\mu}}
[{\bd b}^{\dag}_{\bf a} {\bd d}^{\dag}_ {\bar {\bf b} }  ]^{ \bar \Gamma}_{\bar \mu} \right\}\nonumber\\
&=&\sum_{{\bf a},\mu_{\bf a}}\sum_{{\bf b},\mu_{\bf b}}
\left\{ \left(X^n_{{\bf a}{\bf b};\Gamma}\right)^*
\langle {\bf a} \mu_{\bf a}, \bar {\bf b} \bar \mu_{\bf b}| \Gamma\mu \rangle 
{\bd d}^{\bar {\bf b} \bar \mu_{\bf b}} {\bd b}^{{\bf a} \mu_{\bf a}}
-\left(Y^n_{{\bf a}{\bf b};\Gamma}\right)^* (-1)^{\phi_{\Gamma\mu}}
\langle {\bf a} \mu_{\bf a}, \bar {\bf b} \bar \mu_{\bf b}| \bar \Gamma \bar \mu \rangle 
{\bd b}^ \dag _{\bar {\bf a} \mu_{\bf a}} {\bd d}^ \dag_{{\bf b} \bar \mu_{\bf b}}
\right\}
~~~,
\label{eq16}
\eeqa
where ${\bf a}=\pi_a, k_a, j_a, Y_a,T_a$ ($\bar {\bf b}=\pi_b, k_b,
j_b, -Y_b,T_b$) and $\mu_{\bf a}=m_a c_a T_{az}$ ($\bar \mu_{\bf  b}=-m_b \bar c_b -T_{bz}$) 
are the irreducible representations and the third components of
quarks (antiquarks), respectively, while
 $\langle {\bf a} \mu_{\bf a}, \bar {\bf b} \bar \mu_{\bf b}|\Gamma\mu \rangle$ 
is a short-hand notation for the product over all Clebsch-Gordan
coefficients involved (spin, color and flavor).
The quantities $X^n_{{\bf a}{\bf b};\Gamma}$ are the {\it  forward-going-amplitudes} and
$Y^n_{{\bf a}{\bf b};\Gamma}$ are the {\it backward-going-amplitudes}.
The TDA limit is obtained when the backward-going-amplitudes $Y^n_{{\bf a}{\bf b};\Gamma}$ are
set to zero. All the details about the phonon operators and the implementation of the TDA
and RPA methods are presented in Appendix \ref{AP:TDA-and-RPA}.

\bigskip
Before discussing our TDA and RPA results, we describe our general procedure
for the numerical calculations: \\
\begin{itemize}

\item We are not considering dynamical gluons, but their main effects are accounted for
by an effective confining potential, obtained from a self-consistent
treatment (Dyson-Schwinger equations) of the Yang-Mills degrees of
freedom in the presence of static quarks \cite{Adam2001}. 

\item For the fermion fields, we implement the canonical quantization 
in terms of a basis of wave functions and the corresponding creation (annihilation)
operators, see Eq.\ (\ref{fermion-field}). 
In particular, we have chosen the harmonic oscillator basis for the
expansion of the fermion fields and used it to  exactly diagonalize the
kinetic Dirac term (see Section \ref{Sec:Kinetic-term}) in each
flavor-isospin sector, in order to obtain effective quarks and antiquarks.

\item Then, the effective QCD Hamiltonian has a diagonal single-particle term representing the  kinetic energies (including the masses) of
the effective quarks (Eq.\ (\ref{eq9})), while now the matrix elements of the color Coulomb confining interaction
${\bd H}_{\text{Coul}}$ are not only spin and color dependent but also flavor-isospin dependent, as shown in Eq. (\ref{eq-new-matrix-elements}),
as a consequence of the unitary transformation $\alpha^{j,T}_{\tau,(N l); \lambda \pi, k}$. 

\item  It is reasonable to expect that the masses of the effective
up, down and strange quarks are higher than the corresponding current
quark masses, and different from each other. These differences
account for the effects of the chiral and flavor symmetry breaking
and they are represented by the unitary transformation of Eq. (\ref{eq3}).
Here, we consider that the current strange quarks have a higher mass than
the up and down quarks. For simplicity, we use 
up and down quarks with the same mass, {\it i.e.},  $m_s > m_{u,d}$. 
We are then in a position to describe the interaction of the
effective up, down and strange quarks at low energies.

\item In order to study the predictive power of our approach, 
the numerical calculations are restricted first to a particle-hole space 
by means of the TDA where the $1p-1h$ correlations are only taken into account in the 
excited states, keeping the ground state unchanged.
This allows us to give a quantitative analysis of the ground state correlations once the RPA 
method is implemented. The collective TDA eigenstates are compared to experimental data 
in the subspaces of spin, parity, color, flavor-hypercharge and isospin
quantum numbers $\{J^P,(0,0)_C,(Y,T)\}$ for masses
up to 1~GeV, with $J^P=0^{\mp},1^-$ and $(Y,T)=(\pm 1,\frac{1}{2}),(0,0), (0,1)$.

\item In a further step,  we include also ground state $p-h$ correlations by means of the
RPA within the same subspaces $\{J^P,(0,0)_C,(Y,T)\}$. 
By exploring the influence of ground state correlations
in a  comparison of the RPA-phonon eigenvalues with
the corresponding TDA eigenvalues and the experimental
data, we obtain insights into the physics around the pion state, 
as well as a look at the controversial scalar mesons ($J^P=0^+$).

\item Finally, to absorb any dependence of the configurational
space on the cut-off $N_{{\rm cut}}$, we have to implement a  renormalization procedure. 
We will require that the observables are weakly depending on the cut-off of the configurational space.  
To achieve this, the effective  particle energies obtained from the
prediagonalization of the kinetic energy term also have to be independent of
the cut-off, while the bare quark masses ($m_{u,d}$ and $m_s$) and the interaction
parameters (couplings $\alpha$ and $\beta$) get renormalized as 
functions of $N_{{\rm cut}}$. The details of the renormalization procedure
are shown in Appendix \ref{Ap-RN}.  

\end{itemize}

\subsection{TDA and RPA results}
\label{sec3}

In what follows, we present our results and compare them
with the results obtained in \cite{Llanes} and with the experimental data \cite{PDG2016}. 
We also compare the TDA and RPA results with each other in order
to explore the influence of the ground state correlations in the
different subspaces $\{J^P,(0,0)_C,(Y,T)\}$ where both methods are implemented. 
From now on, we are going to omit the singlet color 
notation $(0,0)_C$ from the TDA and RPA subspaces of diagonalization, since 
by construction all the collective TDA and RPA states presented here are color singlet.

In order to grasp the main features of the low-energy sector of the meson spectrum,
it is worth to mention the isospin, spin and parity $T(J^P) $ quantum numbers 
of the states up to about $1~\mbox{GeV}$ \cite{PDG2016}.  
For the pseudoscalar mesons: the pion state has $T(J^P)=1(0^-)$, the charged $K^{\pm}$ and
neutral kaons $K^0$ have $T(J^P)=\frac{1}{2}(0^-)$, and the $\eta, \eta'$ mesons 
have $T(J^P)=0(0^-)$. For the vector mesons: the $\rho$ meson has
$T(J^P)=1(1^-)$, the vector kaon states have
$T(J^P)=\frac{1}{2}(1^-)$, while both the $\omega$ and $\phi$ mesons have  $T(J^P)=0(1^-)$.
For the scalar mesons: the $a_0$ meson has $T(J^P)=1(0^+)$, the
$K^*_0$ has $T(J^P)=\frac{1}{2}(0^+)$, and the $f_0(500), f_0(980)=f^{*}_0$ mesons both have $T(J^P)=0(0^+)$.
There is no pseudovector meson reported below $1~\mbox{GeV}$ \cite{PDG2016}.

The TDA and RPA eigenstates in the
$\{J^P,(Y,T)\}=\{J^P,(0,0)\}$ subspaces show only pure $q\bar q$ ($q=u,d$) and
$s\bar s$ states, {\it i.e.}, no flavor mixing.  The reason for such pure states is the fact that
the effective color-confining interaction is not capable to
produce flavor mixing between pairs, as a consequence of the color 
structure  of the quark (antiquark) color-charge densities $\rho^a$, which are
flavor scalars (see Eq.\ (\ref{eq-new-matrix-elements})). Flavor mixing is
expected to occur, however, in the meson spectrum, {\it e.g.}, for the
$T(J^P)=0,(0^-)$ $\eta$ and $\eta'$ physical states. 
In the present approach, then, the $\eta$ and $\eta'$ states will be
described by pure $q\bar q$ and $s\bar s$ collective TDA and RPA states. 
However, we will explore a flavor mixing procedure, following
\cite{Rujula}, in the subspace $\{0^-,(0,0)\}$, 
{\it i.e.}, for pseudoscalar $\eta,~\eta'$ states. In the case of the
$\omega,\phi$ and $f_0$ mesons, related to the subspaces
$\{1^-,(0,0)\}$ and $\{0^+,(0,0)\}$, respectively, the more common
description of these states in the literature is as pure flavorless states.

For the TDA and RPA results presented here, we have used
 $N_{\text{cut}} = N_{0}=3$ (the renormalization procedure for
higher $N_{\text{cut}}$ is shown in Section \ref{Sec:RN} and Appendix \ref{Ap-RN}). 
In Table \ref{inputs}, we show the values of the
quark masses and interaction parameters used in both calculations. 
For a quantitative evaluation of the importance 
of the Coulomb potential $\frac{-\alpha}{|{\bf x}-{\bf y}|}$
(not considered in \cite{Llanes}) with
respect to the linear potential $\beta |{\bf x}-{\bf y}|$,
we also compare the results of the calculations with $\alpha=0$ to those 
with $\alpha \ne 0$.

For $N_{{\rm cut}}=N_0=3$, we  fit the parameters of the model in such a way that the RPA calculation
reproduces the pion mass (139 MeV) as close as possible, 
while the rest of the RPA spectrum ({\it i.e.}, pseudoscalar, vector and scalar mesons) 
is a consequence of this fit. 
\begin{center}
\begin{table}[h!]
\centering
\begin{tabular}{c | cc cc  cc c c c  c cc cc cc  }
\hline\hline
Set \textbackslash  parameters
&& $m_{u,d} [GeV]$ &&$m_s [GeV]$
&&$\alpha $&&$\beta [GeV^2]$
\\ [0.5ex] \hline
1&& 0.05 &&0.31&& 0.0 && 0.40\\[0.5ex] 
2&& 0.05 &&0.31&& 0.16 && 0.40\\[0.5ex] 
\hline
\hline  
\end{tabular}
\caption{Parameters $\alpha$ and $\beta$ of the interaction $V(|{\bf  x}-{\bf  y}|)$
and the masses of the u,d and s quarks, used for the TDA and RPA calculations.
We have set the harmonic oscillator length to $1/\sqrt{B_0}=0.75\mbox{fm}$.} 
\vspace{0.2cm}
\label{inputs}
\end{table}
\end{center}

We shall now briefly comment on the determination of the C-parity of the states under consideration. 
Given that we restrict our calculations to the $j=\frac{1}{2}$ sector of
the basis, the orbital angular momentum for each quark can only be zero or one. 
Thus, in the relative motion (see Appendix \ref{appB}),
the two orbital angular momenta can only couple to the total
relative orbital angular momentum $l_r$ =0, 1 or 2.
The spin $S$ of the quark-antiquark pair can be zero or one, so that the
possible values of the total spin $J$ are
$J=l_r \pm 1$ and $J=l_r$ for $S=1$, and $J=l_r$ for $S$ = 0.
The parity of a state is given by $P=-(-1)^{l_r}$, where the extra minus sign is due to the
relative intrinsic parities of the quark  and antiquark. In particular, when $J$ and $P$ are
defined, one also knows whether $J=l_r$ or $J=l_r \pm 1$.
When one applies the charge conjugation operator to, for example, a $u{\bar u}$ state, one obtains
${\bar u}u$. In order to relate the latter state to the former, one has to exchange the two fermions
(giving a factor $(-1)$), including their positions ($(-1)^{l_r}$) and their spins 
(a factor of $-(-1)^S$). The eigenvalue $C$ is the product of these factors, giving 
$C=(-1)^{l_r+S}$. With this, we can deduce the $C$-parity of the states of interest: \\
i) Pseudoscalars, $J^P = 0^-$: This implies that $l_r$ is even, {\it i.e.}, that $l_r$ is zero, because
for $l_r=2$, and considering that the spin can maximally be one, the value $J=0$ cannot be reached.
Then $S$ has to be zero, too, because $l_r=0$ and $S=1$ would give $J=1$. When $l_r=S=0$, the
$C$-parity is positive. \\
ii) Vector mesons, $J^P = 1^-$: Because of the negative parity, $l_r$
has to be even, {\it i.e.},  $l_r$ is 0 or 2. In both cases, $S$ has to be one in order to get
$J=1$, hence $C=-$.\\
iii) Scalars, $J^P = 0^+$: Because $P$ is positive, $l_r$ has to be odd. It follows that
$l_r=1$ and $S=1$, {\it i.e.}, $C=+$.

\subsection*{Pseudoscalar mesons.}
For the pseudoscalar mesons four diagonalizations are involved, 
one for each subspace $(Y,T)$, however, the sectors
$(\pm1,\frac{1}{2})$ exhibit the expected degeneracy
and the number of diagonalizations can be reduced to three. 
Since we are not making any
distinction at the moment between charged and neutral mesons, we report one pion-
and one kaon-like state, corresponding to the lowest eigenvalue obtained
for $T=1$ and $T=\frac{1}{2}$, respectively. The $\eta$- and $\eta'$-like states in Table
\ref{TAB:ps-no-FM} are pure $q\bar q$ and $s\bar s$ states respectively, {\it
 i.e.}, no flavor mixing \cite{Dudek2011a,Cao} has been implemented.
\begin{center}
\begin{table}[H]
\centering
\begin{tabular}{cc | cc |c cc c cc c cc c cc c cc c cc}
\hline
State     && Exp. && TDA(Set1)  && TDA(Set2)&& RPA(Set1)  && RPA(Set2)
\\\hline
$\pi$    && 139  &&249&&245                         && 145 &&136
\\
K          && 495   && 468  &&465                      && 468 &&465
\\
$\eta$  && 547   && 249  &&245                      && 145 &&136
\\
$\eta'$ && 957    && 688  &&687                       && 670 &&668
\\
\hline  
\end{tabular}
\caption{Experimental pseudoscalar meson masses [MeV] compared to both
TDA and RPA results, for the parameters given in Table \ref{inputs}.} 
\label{TAB:ps-no-FM}
\end{table}
\end{center}

From Table \ref{TAB:ps-no-FM}, it can be seen that the $T=0$ and $T=1$
subspaces are affected by the influence of the ground state
correlations, while the $T=\frac{1}{2}$ subspace remains unchanged.
In particular, the RPA pion-like state is improved with
respect to the TDA one, and is in good agreement with the experimental value.
The reported pseudoscalar kaon mass is well reproduced in this
approach, contrary to the case of Ref.\ \cite{Llanes}, see Figure \ref{Fig:ps-vec}.

In Table \ref{TAB:ps-no-FM}, we show the results for the $\eta$ and $\eta'$ mesons as TDA and RPA
collective states built from pure $q\bar q$ and $s\bar s$ states, respectively.  
As we have explained before, the interactions that are
responsible for the mixing of the $q\bar q$ and $s\bar s$ states are missing
in the effective Hamiltonian (\ref{eq2}). A proper dynamical
implementation of flavor mixing is beyond the purpose of the present work. 
Our pure $q\bar q$ and $s\bar s$ states, associated with the description
of the $\eta$ and $\eta'$ states, share some similarities with the states 
obtained in Ref.\ \cite{Llanes} and are compared with the latter in Figure \ref{Fig:ps-vec}.

For the sake of completeness, we shall add a few more remarks on the
issue of flavor mixing. In the presence of dynamical gluons, a color singlet quark-antiquark pair 
can be virtually annihilated into gluons which subsequently create a new color singlet 
quark-antiquark pair, not necessarily with the same flavor. In full QCD, a $q \bar{q}$ 
state can thus be converted into a $s \bar{s}$ state and vice versa. The same
mechanism also produces $q \bar{q}$ states from $q \bar{q}$ states 
and $s \bar{s}$ states from $s \bar{s}$ states. 
The amplitudes for these virtual annihilation processes are approximately flavor-independent.

An effective description of flavor mixing in the subspace of the $q \bar{q}$ and $s \bar{s}$ states has
been suggested a long time ago in Ref.\ \cite{Rujula}.
In the present context, we would only include the first $q\bar q$ and
$s\bar s$ states of the TDA or RPA solutions and consider them as a
type of effective degrees of freedom. 
Notice that by doing so, we are clearly imposing a severe restriction, 
because  if we preserve the TDA and RPA mappings from the quark-antiquark basis 
to the phonon basis, any sort of added interaction should affect all 
the solutions and not only the $\eta$- and $\eta'$-like solutions.

In its simplest form, the effective Hamiltonian restricted to the subspace of the
(lowest-lying) $q \bar{q}$ and $s \bar{s}$ states
is given by the $2 \times 2$ matrix \cite{Llanes,Rujula} 
\beqa\label{flavor-mix}
\left( \begin{array}{cc}
M_{q \bar{q}} + 2 H_{FM} & \sqrt{2} \, H_{FM} \\
\sqrt{2} \, H_{FM} & M_{s \bar{s}} + H_{FM} 
\end{array} \right) \,.
\eeqa
The masses $M_{q \bar{q}}$ and $M_{s \bar{s}}$ in (\ref{flavor-mix}) correspond to the
lowest, TDA or RPA, $ q \bar{q} $ and $ s \bar{s} $ eigenvalues of
the Hamiltonian (2), in the subspace $\{0^-,(0,0)\}$. Their values
are shown in Table \ref{TAB:ps-no-FM}.
The parameter $H_{FM}$ represents the flavor-independent amplitude of the virtual annihilation
processes and will simply be adjusted to the data.
The factors of two and $\sqrt{2}$ multiplying $H_{FM}$ are a consequence of the fact
that we do not distinguish between $u \bar{u}$ and $d \bar{d}$
states in the present formulation. In this context, the $q \bar{q}$ state 
is to be identified with the isospin singlet state  
$  | q \bar q \rangle = \frac{1}{\sqrt{2}} \big( | u \bar u \rangle + | d \bar d  \rangle \big) $.
The eigenvalues of the flavor mixing matrix (\ref{flavor-mix}) are given by
\beqa
E_\pm=\frac{1}{2} 
\left(  M_{q\bar q}+M_{s\bar s}+3H_{FM}  \pm \sqrt{9H_{FM}^2
+2H_{FM}(M_{q\bar q}-M_{s\bar s})+(M_{q\bar q}-M_{s\bar s})^2 } \right) ~.
\eeqa

In Table \ref{tab:flavor-mix}, we show the fit to the experimental masses, for both TDA and RPA frameworks,
and the corresponding values of the phenomenological parameter $H_{FM}$.
\begin{center}
\begin{table}[H]
\centering
\begin{tabular}{cc | cc c cc c cc c cc c cc c cc c cc}
\hline
\hline
State     && Exp. &&TDA(Set2) && $H_{FM}$&& RPA(Set2) && $H_{FM}$
\\\hline
$\eta$     && 547  &&436 && 154&& 356 && 170
\\
$\eta'$  && 957  &&958 && 154&& 958 && 170
\\
\hline  
\end{tabular}
\caption{Experimental and flavor-mixed TDA and RPA results, for the $\eta$ and $\eta'$ masses [MeV].}
\label{tab:flavor-mix}
\end{table}
\end{center}

For a complete dynamical description of the flavor mixing processes 
in the Coulomb gauge formalism, dynamical gluon degrees of
freedom would have to be included in the effective Hamiltonian
(see Eq.\ (\ref{eq1})) to allow for transitions between quark-antiquark and
multi-gluon states. Such an extension is left for future work.

\subsection*{Vector mesons.} 
For the vector mesons calculated with the masses and coupling constants 
listed in Table \ref{inputs}, we have obtained the TDA and RPA results shown in Table \ref{vec1}.
\begin{center}
\begin{table}[H]
\centering
\begin{tabular}{cc | cc c cc c cc c cc c cc c cc c cc}
\hline
\hline
State     && Exp. && TDA(Set1)  && TDA(Set2)&&RPA(Set1)  && RPA(Set2)
\\\hline
$\rho$        && 770  &&688&&687         && 677 &&676
\\
$\omega$  && 782   &&688&&687        && 677 &&676
\\
$K^*$         && 892   &&941&&942        && 941 &&942
\\
$\phi$       && 1020 &&1144&&1147       && 1142 &&1145
\\
\hline  
\end{tabular}
\caption{Experimental vector meson masses [MeV] compared to both
TDA and RPA results, for the parameters given in Table \ref{inputs}.}
\label{vec1}
\end{table}
\end{center}
The TDA and RPA eigenvalues are in good correspondence with the experimental data.
The energy differences between the experimental vector masses and our
TDA and RPA results are at about $5-12\%$. When assessing the success of
the method, one should take the simplicity of the
procedure and the low computational cost into account.

In Table \ref{vec1}, we show the results for the $\omega$ and $\phi$ mesons as 
TDA and RPA collective states built by pure $q\bar q$ and $s\bar s$
states, respectively. The latter is the most common description for
these states in the literature. In \cite{Bramon}, a flavor mixing
mechanism was used to describe these states, the resulting mixing angle
was very small, in agreement with the picture of pure states. 
In the present approach the effects of an additional interaction,
similar to that described for the pseudoscalar states
could yield a better agreement with the $\rho$ and $\phi$
experimental masses, but should be small.

\begin{figure}[H]
\begin{center}
\includegraphics[width=14.8cm,height=10.0cm]{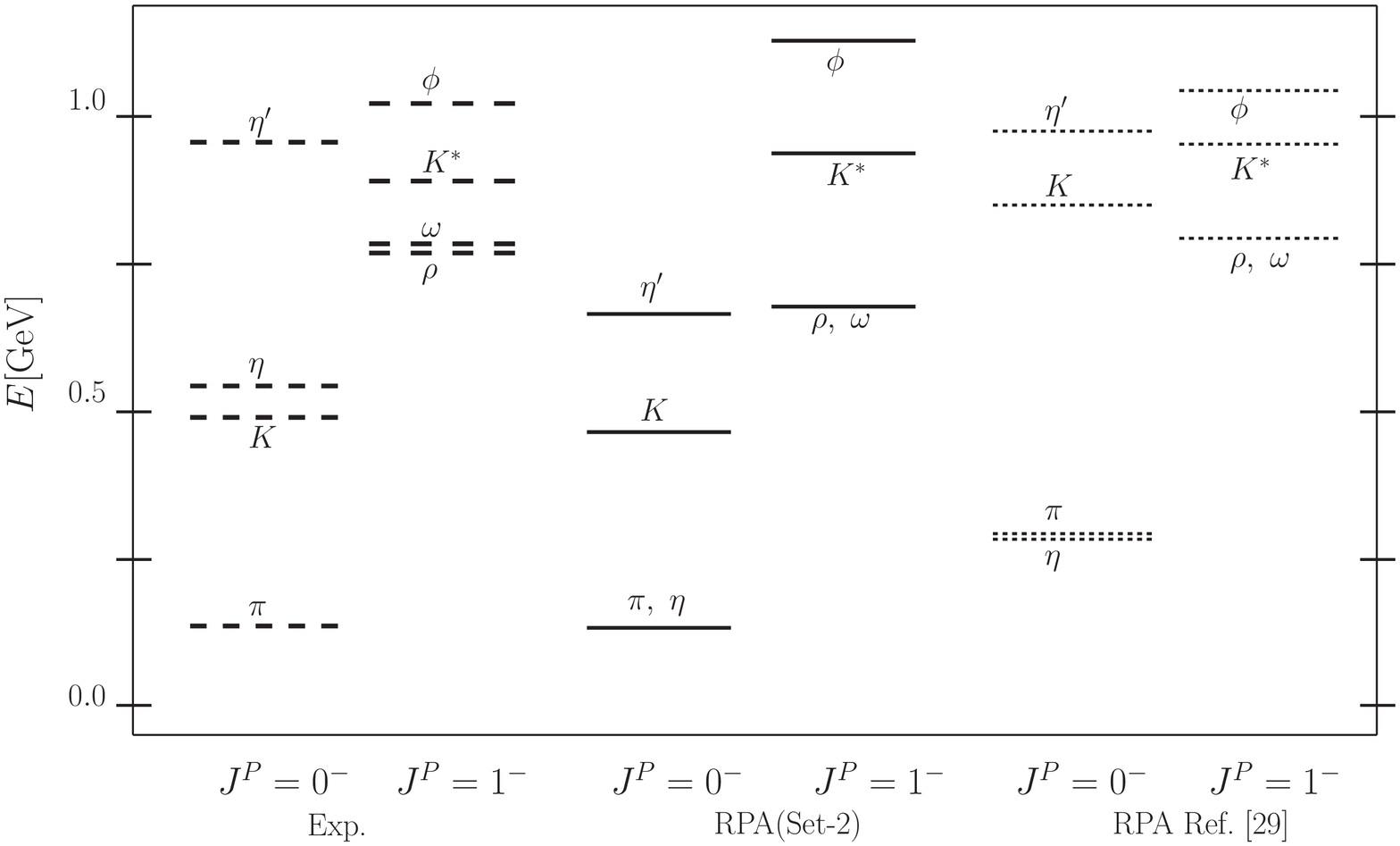}
\caption{RPA pseudoscalar and vector results of Set-2 compared with
those obtained in Ref. \cite{Llanes} and with experimental data.}
\label{Fig:ps-vec}
\end{center}
\end{figure}

In Figure \ref{Fig:ps-vec}, we show the RPA results obtained with
the Set-2 of parameters, {\it i.e.}, the results that better reproduce the
pion energy, compare them to the non-perturbative approach published in \cite{Llanes} and
to the experimental values \cite{PDG2016}. 
As can be seen from Figure \ref{Fig:ps-vec}, the experimental mass of the pseudoscalar kaon is well
reproduced in the present approach, and much less satisfactorily so
in \cite{Llanes}. In general, our results for the kaon-like states, for
$J^P=0^-$, $1^-$ (and $0^+$, see below), are
in good correspondence with the experimental values, once the pion-like state is fitted 

It is apparent from the results shown in Figure \ref{Fig:ps-vec} that
several features of the pseudoscalar and vector meson spectrum are reproduced, 
like those associated with the spin and flavor-isospin dependence, 
see also Tables \ref{TAB:ps-no-FM} and \ref{vec1}.
The energy difference between vector and pseudoscalar 
mesons with the same flavor-isospin (quark content) but different spin, 
like $\rho$ and $\pi$ or the $K^*$ and $K$ states, are reproduced
satisfactorily, while the energy difference between mesons with the
same spin but different flavor-isospin, like $K$ and $\pi$ or
$K^*$ and $\rho$, are reproduced acceptably.
These spin and flavor-isospin dependence effects 
seem to be hidden in the QCD Hamiltonian of  Eq.\ (\ref{eq1}) as well as in
the effective color-confining interaction of Eq.\ (\ref{Eff-QCD-H}).
The main differences between our calculated spectrum and the experimental one are in the
$\eta$ and $\eta'$ sector.

\subsection*{Scalar mesons.}
We have performed the same analysis as for the pseudoscalar
and vector states also for the scalar states.
The scalar mesons below $1~\mbox{GeV}$ are very controversial
\cite{PDG2016,Stone,EichmannFischer}, given that 
the scalar resonances are difficult to identify because some of them
have large decay widths, which cause a strong overlap between
resonances and background.
However, a few scalar mesons are listed in the PDG \cite{PDG2016}.
Their quantum numbers $T(J^P)$ and widths are:  for the quantum numbers
$\frac{1}{2}(0^+)$, the $K^*_0(800)$  (or $\kappa$) with a width of $547~\mbox{MeV}$; for
 $1(0^+)$, the $a_0(980)$ with a width of $50-100~\mbox{MeV}$; and 
for  $0(0^+)$, the $f_0(500)$ (or $\sigma$) with a width of $400-700~\mbox{MeV}$ 
and the $f_0(980)$ with a width of $10-100~\mbox{MeV}$. 

In Table \ref{sc1}, we present the scalar TDA and RPA results with masses
approximately 1 GeV. In the subspace with $(Y,T)=(0,0)$, the $f_0(500)$ is the first
TDA and RPA predicted pure $q\bar q$ state, while the 
$f_0(980)$ is the first predicted pure $s\bar s$ state. 
In \cite{Oller}, the possibility that the $f_0(500)$  and
$f_0(980)$ were related was analized.  The reported mixing angle was about $19^{\circ}$. 
This picture is not that far from the pure $q\bar q$ and
$s\bar s$ description presented in Table \ref{sc1}.  However, our TDA
and RPA results indicate that a sizeable flavor mixing would be needed
to arrive at a satisfactory description of the experimental $f_0$ masses.

For the case of states with $T=\frac{1}{2}$ and $T=1$ {\it i.e.}, $K_0^*$ and $a_0$, respectively, 
we show in Table \ref{sc1} the TDA and RPA results which are closest to the experimental values. 
The other states obtained in these subspaces with energies up to 1 GeV are discussed below.
\begin{center}
\begin{table}[H]
\centering
\begin{tabular}{cc | cc c cc c cc c cc c cc c cc c cc}
\hline
State     && Average && TDA(Set1)  && TDA(Set2)&& RPA(Set1)  && RPA(Set2)
\\\hline
$f_0(500)$   && (400-500)  &&259&&256&&154&&145
\\
$K_0^*(800)$ && 682          &&789&&788&&789&&788
\\
$f_0(980)$  &&  990           &&701&&700&&684&&683
\\
$a_0(980)$   && 980 &&903,1001&&910,1005&&870,961&&877,965
\\
\hline  
\end{tabular}
\caption{Experimental scalar meson masses in units of [MeV] \cite{PDG2016}, compared to both
TDA and RPA results, for the Set-2 of parameters given in Table \ref{inputs}.
The average values are also taken from \cite{PDG2016}.} 
\label{sc1}
\end{table}
\end{center}

The average values given in Table \ref{sc1} account for the large number
of results reported in the literature.

It  has been argued that the
$T(J^{P })=0(0^{+})$-states $f_0(500)$ and $f_0(980)$ have some similarities  
to the $T(J^{P})=0(0^{-})$-states $\eta(547)$ and $\eta'(957)$.  
Particularly, their reported masses are almost the same. 
In the present approach, 
although the TDA and RPA eigenvalues are lower than the experimental values, 
similarities in the masses between the pseudoscalar and scalar states
belonging to the subspaces $(Y,T)=(0,0)$ are also observed
in Tables \ref{TAB:ps-no-FM} and \ref{sc1}.

In \cite{Llanes}, the reported TDA and RPA energy differences between the $f_0$
states are $447~\mbox{MeV}$ and $513~\mbox{MeV}$, respectively. In the
present approach, the same differences, for the Set-2 of parameters of
Table \ref{inputs}, are $444~\mbox{MeV}$ and $538~\mbox{MeV}$, respectively. 
These results are in good agreement with the energy difference between 
the experimental $f_0$ masses, which is $480~\mbox{MeV}$.
\begin{figure}[H]
\begin{center}
\includegraphics[width=14.8cm,height=10.0cm]{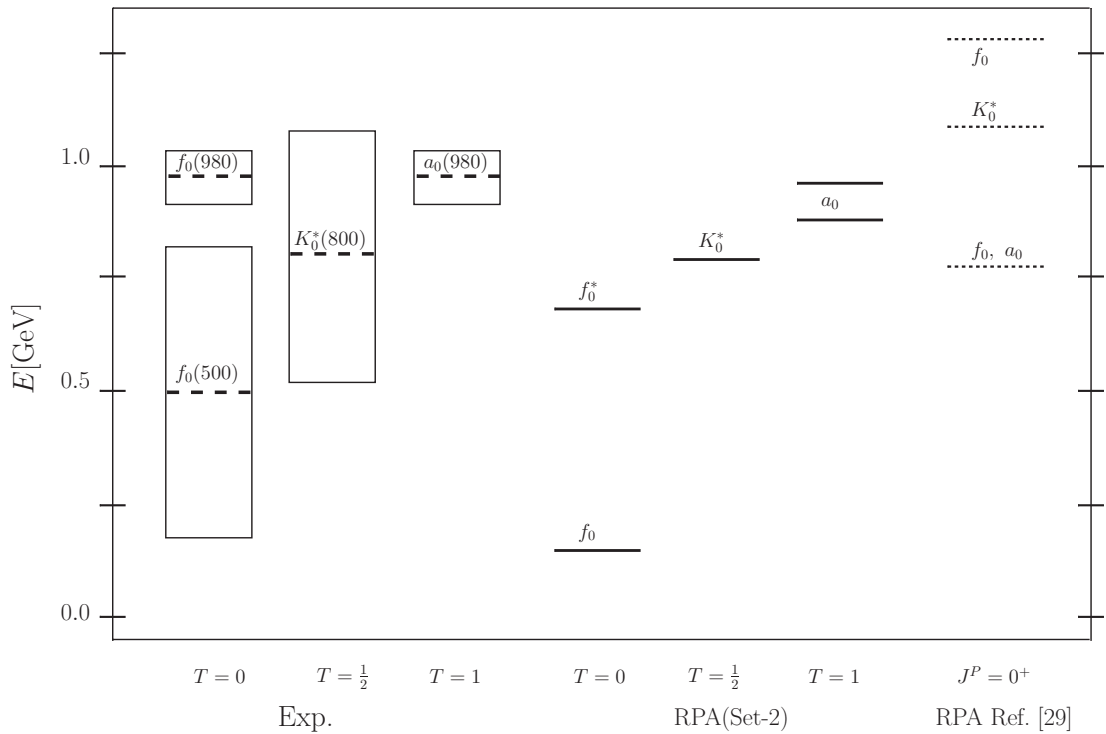}
\caption{Results for scalar mesons of Set-2 in the RPA framework, compared with experimental data.}
\label{Fig:sc}
\end{center}
\end{figure}

The RPA results for scalar mesons obtained in Table \ref{sc1} for Set-2
of Table \ref{inputs}, are compared in Figure \ref{Fig:sc} with the 
scalar states reported in Ref. \cite{Llanes} and with the experimental values.
In our approach, the clearest identification can be made
in the sector with $(Y,T)=(\pm1, \frac{1}{2})$, {\it i.e.}, for the
scalar kaon states. For the $K_0^*(800)$ or $\kappa$ meson, 
several experimental data are reported 
\cite{PDG2016}.  The theoretical predictions for this state
range from $594~\mbox{MeV}$ to $905~\mbox{MeV}$, while the experimental values
locate it between $706$ and $855~\mbox{MeV}$. In our TDA and RPA calculations,
we find one state at about $800~\mbox{MeV}$. 
Finally, for the scalar meson with $(Y,T)=(0,1)$, we find two states
that fit very well with the $a_0(980)$, however, it is hard to say 
at this time whether the physical $a_0$ state could correspond to one of these states,
or even a mixture of them.

As we have mentioned before, one $q {\bar q}$ state and one $s {\bar s}$
state have been used to describe the $f_0$- and $f_0^*$-like states, respectively.
As can be seen from Figure \ref{Fig:sc}, the results obtained in \cite{Llanes}
do not show a second $f_0$ state below 1~GeV, and their
$f_0$ and $a_0$ states are degenerate. In our calculations,
the $f_0$- and $a_0$-like states belong to different
subspaces. The subspace containing the $a_0$-like 
state does not allow $s\bar s$-quark representations.

The understanding of the structure of the rather controversial scalar
states requires, in addition, the calculation of their widths, something that
is beyond the scope of the present work.
Such a calculation will also provide some insights about
one state that appears in the $(Y,T)=(\pm1, \frac{1}{2})$-sector
of the TDA and RPA results at about $480~\mbox{MeV}$, and two states in
the $(Y,T)=(0,1)$-sector located at about $200$ and
$550~\mbox{MeV}$, respectively. These states could be
background states to those reported in Table \ref{sc1}, which 
may  lead to a very broad state rather than a sharp one.

\subsection{Renormalization of the parameters.}
\label{Sec:RN}
In this section we briefly discuss the numerical renormalization of the 
masses and interaction strengths. As we mentioned in Section
\ref{Sec:Kinetic-term}, in order to maintain the observables unchanged, 
we need to renormalize the quark masses and the
couplings of the Hamiltonian when the cut-off ($N_{{\rm cut}}$) of the harmonic
oscillator basis is increased. In Appendix
\ref{Ap-RN}, we explain the philosophy of the renormalization and the
ansatz  used to renormalize the quark masses and the couplings in more detail.

In Table \ref{RN-parameters}, we show the variation of
the masses and interaction parameters with the cut-off that maintains
the single-particle energies, ($\varepsilon_{k,j,Y,T}$ with $j=\frac{1}{2}$) for
each flavor-isospin, {\it i.e.}, $m_q^{eff}=\varepsilon_{1,\frac{1}{2},\pm\frac{1}{2},\frac{1}{2}}$ and
$m_s^{eff}=\varepsilon_{1,\frac{1}{2},0,0}$, invariant. As a result,
the RPA spectrum also remains approximately invariant.
\begin{center}
\begin{table}[h!]
\centering
\begin{tabular}{c  cc cc  cc c c c  c cc cc cc  }
\hline\hline
$N_{{\rm cut}}$ && $m_{q} [GeV]$ &&$m_s [GeV]$ &&$\alpha $&&$\beta [GeV^2]$
&& $m^{eff}_{q} [GeV]$ &&$m^{eff}_s [GeV]$ &&$\pi_{RPA} [GeV]$
\\ [0.5ex] \hline
3&& 0.050 &&0.310 && 0.16 && 0.40
&&0.258 && 0.400 && 0.136
\\[0.5ex] 
5&& 0.142 &&0.337 && 0.23 && 0.28
&&0.258 && 0.400 && 0.136
\\[0.5ex]
7&& 0.173 &&0.351 && 0.29 && 0.22
&&0.258 && 0.400 && 0.137
\\[0.5ex] 
9&& 0.191 &&0.360 && 0.33 && 0.19
&&0.258 && 0.400 && 0.138
\\[0.5ex]
11&& 0.202 &&0.367 && 0.38 && 0.17
&&0.258 && 0.400 && 0.138
\\[0.5ex] 
13&& 0.210 &&0.371 && 0.42 && 0.15
&&0.258 && 0.400 && 0.137
\\[0.5ex] 
\hline
\hline  
\end{tabular}
\caption{Numerical renormalization of the masses and interaction parameters.} 
\vspace{0.2cm}
\label{RN-parameters}
\end{table}
\end{center}

As an example, we show in Table \ref{RN-parameters} the mass of the
RPA pion-like state as a function of the cut-off. The mass of this state remains
approximately constant, while the quark masses and the couplings change according to
\beqa
m^{f}_{N_{{\rm cut}}} & = & 
m^{f}_{N_0} \sqrt{x^f_{{\rm N}_{{\rm cut}}}}
\label{Eeq3c}
\eeqa
and
\beqa
\alpha_{N_{{\rm cut}}} & = & \alpha_{N_0}  \sqrt{x_{N_{{\rm cut}}}}
\nonumber \\
\beta_{N_{{\rm cut}}} & = & \beta_{N_0} / \sqrt{x_{N_{{\rm cut}}}}
\label{Eeq3d}
\eeqa
with $x_{N_{{\rm cut}}}  \approx  1 + t_1 \left(N_{{\rm cut}}-N_0\right)^{t_2} $
and $x^f_{N_{{\rm cut}}}  \approx  1 + t_1^f \left(N_{{\rm cut}}-N_0\right)^{t_2^f}$, respectively. 
The superscript $f=q,s$ distinguishes between the up, down ($q=u=d$)
and strange ($s$)  quarks.

\begin{figure}[H]
\begin{center}
\includegraphics[width=12.8cm,height=8.0cm]{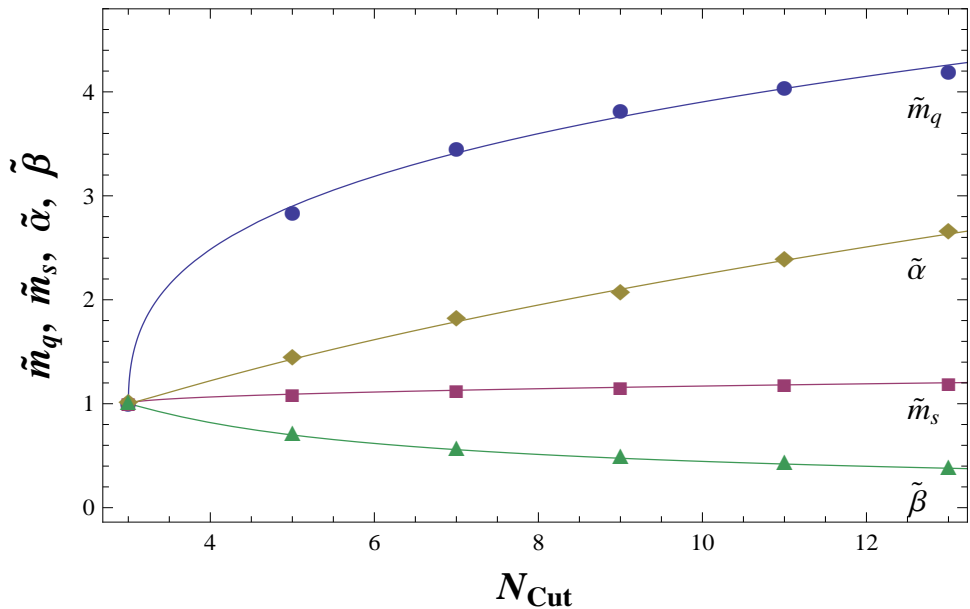}
\caption{Renormalization of the masses and interaction constants as a function of $N_{{\rm cut}}$.}
\label{RN-plots}
\end{center}
\end{figure}

In Figure \ref{RN-plots}, we show the dimensionless values 
$\tilde m_q=\frac{m^{N_{\rm cut}}_q}{m^{N_0}_q}$, 
$\tilde m_s=\frac{m^{N_{\rm cut}}_s}{m^{N_0}_s}$,
$\tilde \alpha=\frac{\alpha_{N_{\rm cut}}}{\alpha_{N_0}}$
and $\tilde \beta=\frac{\beta_{N_{\rm cut}}}{\beta_{N_0}}$ as
functions of the cut-off.

The values for the parameters $t_1,t_2,t_1^f,t_2^f$ are listed in Table \ref{RN-fits}.
\begin{center}
\begin{table}[h!]
\centering
\begin{tabular}{c  cc cc  cc c c c  c cc cc cc  }
\hline\hline
$t_1$ && $t_2$ &&$t_1^q$ &&$t_2^q $&&$t_1^s$&& $t_2^s$ 
\\ [0.5ex] \hline
0.49&& 1.08 &&5.17 && 0.52 && 0.13 &&0.52
\\[0.5ex] 
\hline
\hline  
\end{tabular}
\caption{Numerical fits for the parameters of the renormalization functions in Eqs.\ (\ref{Eeq3c}) and (\ref{Eeq3d}).} 
\vspace{0.2cm}
\label{RN-fits}
\end{table}
\end{center}

Based on the results of Table \ref{RN-fits}, the renormalization of the parameters can be written in the simpler form
\beqa
m^{f}_{N_{{\rm cut}}} & = & 
m^{f}_{N_0} \sqrt{  1 + t_1^f (N_{{\rm cut}}-N_0)^{\frac{1}{2}}}\nonumber\\
\alpha_{N_{{\rm cut}}} & = & \alpha_{N_0}  \sqrt{ 1 + \frac{1}{2} \left(N_{{\rm cut}}-N_0\right)  }
\nonumber\\
\beta_{N_{{\rm cut}}} & = & \frac{\beta_{N_0}}
{  \sqrt{ 1 + \frac{1}{2} \left(N_{{\rm cut}}-N_0\right)  }}
~~.
\eeqa

\section{Conclusions}
\label{Conclusions}
We have applied standard many-body methods, adapted from the nuclear
many-body problem, to a simplified effective QCD Hamiltonian suitable for the
description of low-lying mesons.
The interaction between color-charge densities was approximated by a
combination of a Coulomb plus linear confining potential, which takes the gluon
contribution into account in an effective way.
We have used the  harmonic oscillator wave functions as a basis for the
quantization of the fermion fields. It was shown that this basis is
quite convenient  for the calculations.

A prediagonalization of the kinetic energy term was performed,
which provided a basis where effective quarks and antiquarks
operators could be defined. 
We deduced all the expressions for the
QCD-motivated Hamiltonian in this effective basis. 
Then, the Hamiltonian was diagonalized in the basis 
of quark and antiquark pairs, by means of both the 
{\it Tamm-Dancoff Approximation} (TDA) 
and the {\it Random Phase Approximation} (RPA). 
The building blocks in this part of the calculation are effective quarks and antiquarks
coupled to color singlets.
The meson spectrum below 1 GeV  was calculated
for configurations belonging to subspaces with quantum numbers $\{J^P,(Y,T)\}$
of pseudoscalar, vector and scalar states. 
In the subspaces with $(Y,T)=(0,0)$, the simplified version of
the color confining interaction used in this approach, was not
able to generate dynamical flavor mixing.

For the meson-like states we obtain the following results:\\
i) The pion-like state of the RPA scheme, 
when the parameter $\alpha$  of the interaction
is different from zero, is in better agreement with the experimental
value than the TDA one, for the same parametrization. 
This is due to the ground state correlations present in
the RPA framework.\\
ii) The kaon-like states are well reproduced. These states are not influenced
by the ground state correlations due to the scalar
structure of the RPA vacuum in the flavor-isospin channel.\\
iii) In this description, we were able to accommodate scalar mesons.
This may contribute to clarify the controversies found in the
literature about these states.\\
iv) Besides the  $\eta$- and $f_0$-like states, which in the RPA scheme 
are quite low with respect to the data, the rest of the
light meson spectrum is well reproduced,
as shown in Figures \ref{Fig:ps-vec} and \ref{Fig:sc}.
The difficulty with the $\eta$- and $f_0$-like states is a clear indication
of the lack of a dynamical flavor
mixing term in the Hamiltonian used here.
To incorporate the corresponding virtual pair annihilation processes in
our approach would require the introduction of dynamical gluonic degrees of
freedom, so that the Hamiltonian could allow for a mixing of quark-antiquark
states with states of at least two dynamical gluons.

In the present approach, we have adjusted the parameters by fitting
the pion-like state within the RPA scheme to the experimental
value. The rest of the spectrum is a consequence of such a fit,
because we have not performed further adjustment of the masses and
couplings appearing in the model.

The renormalization procedure of the parameters in
terms of the cut-off of the  
configurational space, $N_{\rm cut}$, was presented in Section
\ref{Sec:RN} and Appendix \ref{Ap-RN}. We have shown that after the renormalization,
the RPA results are weakly dependent on the cut-off.

The present work shows that the TDA and RPA many-body methods 
can give valuable insight into the non-perturbative regime of QCD. 
The procedure described here 
provides a straightforward manner to calculate
the complete spectrum of hadron states,
and it can be extended towards more complete effective descriptions of QCD.

\section*{Acknowledgements}
One of the authors (T.Y-M) thanks the National Research Council of
Argentina (CONICET) for a post-doctoral scholarships. (O.C.) is a member 
of the scientific career of the CONICET. (P.O.H and D.A.A-Q) acknowledges financial help from
DGAPA-PAPIIT (IN100315) and from CONACYT (Mexico, grant 251817). 
This work has been supported financially
by the CONICET (PIP-282) and by the ANPCYT. D.A.A-Q. acknowledges a scholarship received
from CONACyT. A.W. is grateful to the Instituto de Ciencias Nucleares of
Universidad Nacional Aut\'onoma de M\'exico for the warm hospitality extended
to him during a five-months stay in 2016. He also acknowledges support
by Conacyt project no.\ CB-2013/222812 and by CIC-UMSNH.

\appendix 
\section{The kinetic energy and the effective basis}
\label{appA}
The kinetic energy in terms of the creation and annihilation operators in the harmonic
oscillator basis is given by \cite{Yepez2010}
\beqa
{\bd K} & = &
\sum_{j\tau_i N_i l_i} 
\sum_{m c f} 
K^{j,T}_{\tau_1(N_1l_1),\tau_2(N_2l_2)}
{\bd q}^\dagger_{\tau_1(N_1 l_1)jmcf}  {\bd q}^{\tau_2(N_2 l_2)jmcf}
~~~,
\label{Aeq3}
\eeqa
where the matrix elements are 
\beqa
K^{j,T}_{\tau_1(N_1l_1),\tau_2(N_2l_2)}
= \left\{
\begin{array}{cc}
k^j_{N_1N_2} & {\rm if}~l_1=j+\frac{1}{2} ~,~ l_2=j-\frac{1}{2} 
~,~ \tau_1~\ne~\tau_2 \\
k^j_{N_2N_1} & {\rm if}~l_1=j-\frac{1}{2} ~,~ l_2=j+\frac{1}{2} 
~,~ \tau_1~\ne~\tau_2 \\
m^T_0 & {\rm if} ~(N_1l_1)=(N_2l_2) ~,~ \tau_1=\tau_2=+\frac{1}{2} \\
-m^T_0 & {\rm if} ~(N_1l_1)=(N_2l_2) ~,~ \tau_1=\tau_2=-\frac{1}{2}
\\
0 & {\rm in}~{\rm all}~{\rm other}~{\rm cases} 
\end{array}
\right.~~~,
\label{Aeq4}
\eeqa
and $k^j_{NN^\prime}$ is defined as \cite{Yepez2010}
\beqa
k^j_{NN^\prime} 
& = & \sqrt{B_0} \left( \sqrt{\frac{N-j+\frac{3}{2}}{2}} \delta_{N^\prime,N+1}
+  \sqrt{\frac{N+j+\frac{3}{2}}{2}} \delta_{N^\prime,N-1} \right)
~~~.
\label{Aeq2}
\eeqa

The diagonalization of the kinetic energy  
implies a transformation to a new basis of creation operators 
\beqa
{\bd q}^\dagger_{\tau (N l)jmcf} & = &
\sum_{\lambda \pi k} \left( \alpha^{j,T}_{\tau (N l),\lambda \pi k}\right)^* 
{\bd Q}^\dagger_{\lambda \pi k j mcf} \delta_{\pi, (-1)^{\frac{1}{2}-\tau+l}}
~~~,
\label{ap:old-new}
\eeqa
where we may take the transformation coefficients
$\alpha^{j,T}_{\tau (N l),\lambda \pi k}$ to be real.

After the prediagonalization in  Eq.\ (\ref{eq:prediag}), the association with quarks and antiquarks is,
according to the Dirac picture, 
\beqa
{\bd Q}^\dagger_{\frac{1}{2} \pi kjmcf} & \rightarrow & {\bd b}^\dagger_{\pi kjmcf}
\nonumber \\
{\bd Q}^\dagger_{-\frac{1}{2} \pi kjmcf} & \rightarrow & {\bd d}_{\pi kjmcf}
~~~.
\label{eq4}
\eeqa
The operator ${\bd b}^\dagger_{\pi kjmcf}$ is a quark creation operator and 
${\bd d}_{\pi kjmcf}$ an antiquark annihilation operator with lower indices.

The rule to raise and lower indices is taken from {\cite{new-basis},
\beqa
{\bd b}^{\dagger \pi kjm {\bar c}{\bar f}} & = &
(-1)^{\chi_f} (-1)^{\chi_c+j+m} {\bd b}^\dagger_{\pi kj-mcf}
\nonumber \\
{\bd b}^{\dagger}_{\pi kjm cf} & = &
(-1)^{\chi_f} (-1)^{\chi_c+j-m} {\bd b}^{\dagger \pi kj-m{\bar c}{\bar f}}
\nonumber \\
{\bd b}^{\pi kjm cf} & = &
(-1)^{\chi_f} (-1)^{\chi_c+j-m} {\bd b}_{\pi kj-m{\bar c}{\bar f}}
\nonumber \\
{\bd b}_{\pi kjm {\bar c}{\bar f}} & = &
(-1)^{\chi_f} (-1)^{\chi_c+j+m} {\bd b}^{\pi kj-mcf}
~~~,
\label{eq5}
\eeqa
and similarly,
\beqa
{\bd d}^{\dagger \pi kjm cf} & = &
(-1)^{\chi_f} (-1)^{\chi_c+j+m} {\bd d}^\dagger_{\pi kj-m{\bar c}{\bar f}}
\nonumber \\
{\bd d}^{\dagger}_{\pi kjm {\bar c}{\bar f}} & = &
(-1)^{\chi_f} (-1)^{\chi_c+j-m} {\bd d}^{\dagger \pi kj-m cf}
\nonumber \\
{\bd d}^{\pi kjm {\bar c}{\bar f}} & = &
(-1)^{\chi_f} (-1)^{\chi_c+j-m} {\bd d}_{\pi kj-m cf}
\nonumber \\
{\bd d}_{\pi kjm cf} & = &
(-1)^{\chi_f} (-1)^{\chi_c+j+m} {\bd d}^{\pi kj-m{\bar c}{\bar f}}
~~~.
\label{eq6}
\eeqa

The phase $(-1)^{\chi_f}$ is a short hand notation for 
$(-1)^{\frac{1}{3}+\frac{Y}{2}+T_z}$ \cite{jutta}. The equivalent
notation holds for the color part.

Since we require a unitary transformation which conserves the anti-commutation
relations 
$\left\{ {\bd Q}^{\lambda_1 \pi_1 k_1j_1\mu_1},
{\bd Q}^\dagger_{\lambda_2 \pi_2 k_2j_2\mu_2} \right\}=
\delta_{\lambda_2\lambda_1}\delta_{\pi_2\pi_1}\delta_{j_2j_1}\delta_{k_2k_1}\delta_{\mu_2\mu_1}$
with $\mu_i=m_i c_i f_i $, we have 
\beqa
\left\{ {\bd q}^{\tau_1(N_1 l_1)jm_1 c_1f_1} ,
{\bd q}^\dagger_{\tau_2(N_2 l_2)jm_2 c_2f_2} \right\}& = & 
\delta_{\tau_2\tau_1}
\delta_{N_2N_1} \delta_{l_2l_1} 
\delta_{m_2m_1}\delta_{c_2c_1}\delta_{f_2f_1}
\nonumber \\
& = & 
\sum_{\lambda  \pi k} 
\alpha^{j,T_1}_{\tau_1(N_1l_1),\lambda \pi k} \,
\alpha^{j,T_2}_{\tau_2(N_2l_2),\lambda \pi  k} \,
\delta_{m_2m_1} \delta_{c_2c_1} \delta_{f_2f_1}
~~~,
\label{eq7}
\eeqa
where $\delta_{f_2 f_1}$ is a short notation for $\delta_{Y_2 Y_1} \delta_{T_2 T_1} \delta_{T_{2z} T_{1z}} $.
Thus the matrix $\alpha$ is unitary and satisfies the orthogonality relation
\beqa
\sum_{\lambda \pi k} \alpha^{j,T}_{\tau_1(N_1l_1),\lambda \pi k} \,
\alpha^{j,T}_{\tau_2(N_2l_2),\lambda \pi k}
& = &  \delta_{\tau_2\tau_1}\delta_{N_2N_1}\delta_{l_2l_1} ~~~.
\label{eq8}
\eeqa

\section{The Coulomb-Hamiltonian}
\label{appB}
The Coulomb Hamiltonian in (\ref{eq2}) is written in terms of the fermion
creation and annihilation operators of   Eq. (\ref{fermion-field}). 
The SU(3) color generators $T_C$ and $T^C$ in Eq. (\ref{eq2}) are rewritten
in terms of SU(3) Clebsch-Gordan coefficients \cite{jutta}. Starting from 
$\left( T_C\right)^{c_1}_{~~c_2}$ = $\langle (1,0)c_1\mid T_C \mid
(1,0) c_2\rangle$ and applying the Wigner Eckart Theorem, we obtain

\beqa
\left( T_C\right)^{c_1}_{~~c_2} & = &
\left(-1\right)^{\chi_{c_2} + 1} \frac{1}{\sqrt{2}} 
\langle (1,0) c_1, (0,1) {\bar c}_2 \mid (1,1) C \rangle_1
~~~.
\label{Beq0}
\eeqa

The Coulomb Hamiltonian then takes the form
\beqa
{\bd H}_{{\rm Coul}} &=& 
-\frac{1}{2}
\sum_C \sum_{\tau_iN_il_im_{l_i}j_i m_i\sigma_if_ic_i} 
(-1)^{\chi_C}  
\delta_{\tau_1 \tau_2} \delta_{\tau_3 \tau_4}
\delta_{\sigma_1 \sigma_2} \delta_{\sigma_3 \sigma_4}
\delta_{f_1 f_2} \delta_{f_3 f_4}
\nonumber \\
&\times&
\left[ \langle l_1 m_{l_1},\frac{1}{2}\sigma_1\mid j_1 m_1 \rangle
         \langle l_2 m_{l_2},\frac{1}{2}\sigma_2\mid j_2 m_2 \rangle
\frac{(-1)^{\chi_{c_2}+1}}{\sqrt{2}}
\langle (10)c_1,(01){\bar c}_2\mid (11) C\rangle_1
{\bd q}^\dagger_{\tau_1(N_1l_1)j_1 m_1c_1f_1}
{\bd q}^{\tau_2(N_2l_2)j_2 m_2c_2f_2}
\right]
\nonumber \\
&\times&
\left[ \langle l_3m_{l_3},\frac{1}{2}\sigma_3\mid j_3 m_3 \rangle
         \langle l_4m_{l_4},\frac{1}{2}\sigma_4\mid j_4 m_4 \rangle
\frac{(-1)^{\chi_{c_4}+1}}{\sqrt{2}}
\langle (10)c_3,(01){\bar c}_4\mid (11) {\bar C}\rangle_1
{\bd q}^\dagger_{\tau_3(N_3l_3)j_3 m_3c_3f_3}
{\bd q}^{\tau_4(N_4l_4)j_4 m_4c_4f_4}
\right]
\nonumber \\
&\times&
\int d{\bf x} d{\bf y} \Psi^*_{N_1l_1m_{l_1}}({\bf x})\Psi_{N_2l_2m_{l_2}}({\bf x})
V(\mid {\bf x} - {\bf y}\mid)
\Psi^*_{N_3l_3m_{l_3}}({\bf y})\Psi_{N_4l_4m_{l_4}}({\bf y})
~~~,
\label{Beq1}
\eeqa
where $\Psi_{N l m_{l}}({\bf x})$ are the three-dimensional harmonic
oscillator wave functions.
In Eq.\ (\ref{Beq1}), the presence of  
$\delta_{\tau_1 \tau_2} \delta_{\tau_3 \tau_4}$,  $ \delta_{\sigma_1 \sigma_2} \delta_{\sigma_3 \sigma_4}$ 
and $\delta_{f_1 f_2} \delta_{f_3 f_4} $ arises from the color
charge-density structure which does not contain any pseudospin, spin or
flavor-isospin dependence.

Let us concentrate for a moment on the integral over ${\bf x}$ and ${\bf y}$.
By recoupling we obtain
\beqa
&&\sum_{J^\prime M^\prime_j J M_J}
\langle l_1m_{l_1},l_3m_{l_3}\mid J^\prime M^\prime_J \rangle 
\langle l_2m_{l_2},l_4m_{l_4}\mid J M_J \rangle
 \nonumber \\
&&\times
\int d{\bf x} d{\bf y} \left[ \Psi^*_{N_1l_1}({\bf x}) 
\otimes \Psi^*_{N_3l_3}({\bf y})\right]^{J^\prime}_{M^\prime_J}
V(\mid {\bf x} - {\bf y}\mid)
\left[ \Psi_{N_2l_2}({\bf x}) \otimes \Psi_{N_4l_4}({\bf y})\right]^{J}_{M_J}
~~~.
\label{Beq2}
\eeqa

Using the Moshinsky brackets (round brackets) \cite{mosh1,mosh2} for
recoupling the integral in Eq. (\ref{Beq2}), we get

\beqa &&
\sum_{N^\prime_r l^\prime_r N_rl_r; N^\prime_R L^\prime_R N_R L_R}
( N^\prime_r l^\prime_r, N^\prime_R L^\prime_R, J^\prime\mid N_1l_1,N_3l_3,J^\prime)
( N_r l_r, N_R L_R, J\mid N_2l_2,N_4l_4,J)
\nonumber \\
&&\times
\int d{\bf r} d{\bf R} \left[ \Psi^*_{N^\prime_RL^\prime_R}({\bf R}) 
\otimes \Psi^*_{N^\prime_rl^\prime_r}({\bf r})\right]^{J^\prime}_{M^\prime_J}
V(\sqrt{2}r)
\left[ \Psi_{N_RL_R}({\bf R}) \otimes \Psi_{N_rl_r}({\bf r})\right]^{J}_{M_J}
~~~,
\label{Beq3}
\eeqa
where ${\bf R}=  \frac{1}{\sqrt{2}}\left({\bf x}+{\bf y}\right)$ refers to the {\it Center of Mass} (CM) 
and ${\bf r}=\frac{1}{\sqrt{2}}\left({\bf x}-{\bf y}\right)$ to the relative  coordinate. 
Since in the last equation the potential is $R$-independent, we
can easily integrate over $R$. Then the integral in Eq.\ (\ref{Beq3}) becomes

\beqa
&&\sum_{M^\prime_R M_R; m^\prime_r m_r}
\langle L^\prime_R M^\prime_R,l^\prime_r m^\prime_r\mid J^\prime M^\prime_J \rangle
\langle L_R M_R,l_r m_r\mid J M_J \rangle
\nonumber\\
&&
\int d{\bf R} \Psi^*_{N^\prime_RL^\prime_RM^\prime_R}({\bf R}) \Psi_{N_RL_RM_R}({\bf R})
\int d{\bf r} \Psi^*_{N^\prime_rl^\prime_rm^\prime_r}({\bf r}) V(\sqrt{2}r)\Psi_{N_rl_rm_r}({\bf r})
\nonumber\\
&&=
\delta_{N^\prime_RN_R} \delta_{L^\prime_RL_R}
\sum_{M_R; m^\prime_r m_r}
\langle L_R M_R,l^\prime_r m^\prime_r\mid J^\prime M^\prime_J \rangle
\langle L_R M_R,l_r m_r\mid J M_J  \rangle
\int d{\bf r} \Psi^*_{N^\prime_rl^\prime_rm^\prime_r}({\bf r})
V(\sqrt{2}r)\Psi_{N_rl_rm_r}({\bf r})
~.~~
\label{Beq5}
\eeqa
Since $V$ is a scalar, we have as an additional constraint that $l^\prime_r=l_r$
and $m^\prime_r=m_r$. 
The integral itself does not depend
on $m_r$, thus we can replace $m_r$ by $l_r$.
With this replacement, the  summation over $M_R, m^\prime_r,m_r$ with the two
Clebsch-Gordan coefficients becomes a summation over $M_R, m_r$ and
represents the orthogonality relation of the Clebsch-Gordan coefficients,
thus leading to $\delta_{J^\prime J}\delta_{M^\prime_JM_J}$.

With all this, equation (\ref{Beq2}) takes the form
\beqa
&&
\sum_{JM_J;N^\prime_rN_rl_r;N_RL_R}
\langle  l_1m_{l_1},l_3m_{l_3}\mid JM_J \rangle  \langle l_2m_{l_2},l_4m_{l_4}\mid JM_J \rangle
\nonumber \\
&&\times~
( N^\prime_rl_r,N_RL_R,J\mid N_1l_1,N_3l_3,J )
( N_rl_r,N_RL_R,J\mid N_2l_2,N_4l_4,J )
\int d{\bf r} \Psi^*_{N^\prime_r l_rl_r}({\bf r})
V(\sqrt{2}r)\Psi_{N_rl_rl_r}({\bf r})
~~.
\label{Beq7}
\eeqa
Using (\ref{Beq7}), the Hamiltonian (\ref{Beq1}) can be written
\beqa
&&
{\bd H}_{{\rm Coul}}
=  -\frac{1}{2} 
\sum_{N_il_im_{l_i} j_i m_i\sigma_i} \sum_{JM_JN^\prime_rN_rl_rN_RL_R}
\delta_{\sigma_1 \sigma_2} \delta_{\sigma_3 \sigma_4}
~\frac{1}{2}
~\langle l_1m_{l_1},\frac{1}{2}\sigma_1\mid j_1 m_1 \rangle
\langle  l_2m_{l_2},\frac{1}{2}\sigma_1\mid j_2 m_2 \rangle
\nonumber \\
&&\times
\langle  l_3m_{l_3},\frac{1}{2}\sigma_3\mid j_3 m_3 \rangle 
\langle  l_4m_{l_4},\frac{1}{2}\sigma_3\mid j_4 m_4 \rangle
\langle  l_1m_{l_1},l_3m_{l_3}\mid JM_J \rangle  \langle  l_2m_{l_2},l_4m_{l_4}\mid JM_J \rangle
\nonumber \\ 
&&\times
( N^\prime_rl_r,N_RL_R,J\mid N_1l_1,N_3l_3,J )  ( N_rl_r,N_RL_R,J\mid N_2l_2,N_4l_4,J )
\int d{\bf r} \Psi^*_{N^\prime_r l_r l_r}({\bf r}) V(\sqrt{2}r)\Psi_{N_rl_rl_r}({\bf r})
\nonumber \\
&&\times
\bigg\{  \sum_C (-1)^{\chi_C}  \sum_{c_i\tau_if_i} 
\delta_{\tau_1 \tau_2} \delta_{\tau_3 \tau_4}~\delta_{f_1 f_2} \delta_{f_3 f_4}
\langle (10)c_1,(01){\bar c}_2\mid (11) C\rangle_1 
\langle (10)c_3,(01){\bar c}_4\mid (11) {\bar C}\rangle_1
~(-1)^{\frac{1}{2}-\tau_2+\frac{1}{2}-\tau_4}  
\nonumber \\
&&\times   (-1)^{j_2-m_2+j_4-m_4}   (-1)^{\chi_{f_2}+\chi_{f_4}} 
~{\bd q}^\dagger_{\tau_1(N_1l_1)j_1 m_1c_1 f_1} 
{\bd q}_{-\tau_2(N_2l_2)j_2-m_2 \bar c_2 \bar f_2}
{\bd q}^\dagger_{\tau_3(N_3l_3)j_3 m_3c_3 f_3}
{\bd q}_{-\tau_4(N_4l_4)j_4-m_4 \bar c_4 \bar f_4}
\bigg\}  ~.
\nonumber\\
\label{Beq8}
\eeqa

The phase $(-1)^{\chi_{c_2}+\chi_{c_4}}$ appearing in Eq.\ (\ref{Beq1}) is
canceled in Eq.\ (\ref{Beq8}) against the same phase resulting from lowering the indices of the
operators. The remaining $(-1)^{\chi_C}$ phase 
is proportional to the SU(3) coupling to color zero, where the factor
$1/\sqrt{8}$ is missing. The contraction over $\tau_i$ 
({\it i.e.}, $(-1)^{\frac{1}{2}-\tau_2}\delta_{\tau_1 \tau_2}$) and $f_i$  
({\it i.e.}, $(-1)^{\chi_{f_2} } \delta_{f_1 f_2}$) also represents a
coupling to zero, without the factor $1/\sqrt{2}$
in the first case and the factor $1/\sqrt{3}$ in the second case. Correcting for
these factors, we have for the expression $\left\{....\right\}$ in Eq.
(\ref{Beq8}), 
\beqa
\bigg\{....\bigg\}&=&
\sum_{LM_L} (-1)^{j_2-m_2+j_4-m_4}(\sqrt{2})^2(\sqrt{3})^2\sqrt{8}
\langle j_1 m_1,j_2-m_2\mid LM_L\rangle \langle
j_3 m_3,j_4-m_4\mid L-M_L \rangle
\frac{(-1)^{L-M_L}}{\sqrt{2L+1}}
\nonumber \\
&\times&
\left[\left[
{\bd q}^\dagger_{(N_1l_1)j_1}
\otimes
{\bd q}_{(N_2l_2)j_2}
\right]^{0L(11)(00)}
\otimes
\left[
{\bd q}^\dagger_{(N_3l_3)j_3}
\otimes
{\bd q}_{(N_4l_4)j_4}
\right]^{0L(11)(00)}
\right]^{00(00)(00)}_{00~0~~0}
~~~.
\label{Beq9}
\eeqa
The intermediate and total couplings are ordered in the following way:
pseudo-spin, angular momentum, color and flavor.

As a result, we can rewrite the Coulomb Hamiltonian (Eq. (\ref{Beq1})) as
\beqa
{\bd H}_{{\rm Coul}}
&=&  -\frac{1}{2}
\sum_{N_il_im_{l_i} j_i m_i\sigma_iLM_L} ~\sum_{JM_JN^\prime_rN_rl_rN_RL_R}
\frac{1}{2}(\sqrt{2})^2(\sqrt{3})^2\sqrt{8} (-1)^{j_2-m_2+j_4-m_4}
\frac{(-1)^{L-M_L}}{\sqrt{2L+1}}
\nonumber \\
&\times&
\langle  l_1m_{l_1},\frac{1}{2}\sigma_1\mid j_1 m_1 \rangle
\langle  l_2m_{l_2},\frac{1}{2}\sigma_1\mid j_2 m_2   \rangle
 \langle j_1m_1,j_2-m_2\mid LM_L  \rangle
\nonumber \\
&\times&
\langle l_3m_{l_3},\frac{1}{2}\sigma_3\mid j_3 m_3  \rangle
\langle l_4m_{l_4},\frac{1}{2}\sigma_3\mid j_4 m_4  \rangle
\langle  j_3 m_3,j_4-m_4\mid L-M_L  \rangle
\nonumber \\
&\times&
\langle  l_1m_{l_1},l_3m_{l_3}\mid JM_J \rangle  \langle  l_2m_{l_2},l_4m_{l_4}\mid JM_J  \rangle
\nonumber \\
&\times&
( N^\prime_rl_r,N_RL_R,J\mid N_1l_1,N_3l_3,J  )
(  N_rl_r,N_RL_R,J\mid N_2l_2,N_4l_4,J  )
\int d{\bf r} \Psi^*_{N^\prime_r l_r l_r}({\bf r})
V(\sqrt{2}r)\Psi_{N_rl_rl_r}({\bf r})
\nonumber \\
&\times&
\left[\left[ {\bd q}^\dagger_{(N_1l_1)j_1} \otimes {\bd q}_{(N_2l_2)j_2} \right]^{0L(11)(00)}
\otimes
\left[ {\bd q}^\dagger_{(N_3l_3)j_3} \otimes {\bd q}_{(N_4l_4)j_4} \right]^{0L(11)(00)}
\right]^{00(00)(00)}_{00~0~~0}
~~~.
\label{Beq10}
\eeqa

Using the well-known relations for the sums of the products of
three and four Clebsch-Gordan coefficients \cite{Varshalovich}, we
arrive at the final expression for the Coulomb interaction,
\beqa
{\bd H}_{{\rm Coul}}
= -\frac{1}{2}  \sum_{N_il_ij_iL}   V_{\{N_i l_i j_i\}}^{L}
\left[
\left[ {\bd q}^\dagger_{(N_1l_1)j_1} \otimes  {\bd q}_{(N_2l_2)j_2} \right]^{0L(11)(00)}
\otimes
\left[ {\bd q}^\dagger_{(N_3l_3)j_3} \otimes  {\bd q}_{(N_4l_4)j_4} \right]^{0L(11)(00)}
\right]^{00(00)(00)}_{00~0~~0}
~~~,
&
\label{Beq14}
\eeqa
with
\beqa
 V_{\{N_i l_i j_i\}}^{L}
 &=& 
\sum_{JN^\prime_rN_rl_rN_RL_R}
3\sqrt{8}\sqrt{(2j_1+1)(2j_2+1)(2j_3+1)(2j_4+1)}\sqrt{2L+1}(2J+1)
\nonumber \\
&\times &
(-1)^{L+j_2+j_4-J+1}
\left\{
\begin{array}{ccc}
l_1 & L & l_2 \\
j_2 & \frac{1}{2} & j_1
\end{array}
\right\}
\left\{
\begin{array}{ccc}
l_3 & L & l_4 \\
j_4 & \frac{1}{2} & j_3
\end{array}
\right\}
\left\{
\begin{array}{ccc}
l_2 & J & l_4 \\
l_3 & L & l_1
\end{array}
\right\}
\nonumber \\
&\times&
(  N^\prime_rl_r,N_RL_R,J\mid N_1l_1,N_3l_3,J  )
(  N_rl_r,N_RL_R,J\mid N_2l_2,N_4l_4,J  )
\int d^3r \Psi^*_{N^\prime_r l_r l_r}(\vec{r}) V(\sqrt{2}r)\Psi_{N_rl_rl_r}(\vec{r})
~~~.\nonumber\\
\label{Beq15}
\eeqa

The integral over $r$ in Eq.\ (\ref{Beq15}) is a standard one for the three-dimensional harmonic oscillator.
Note that this is an {\it analytic expression} which is easy to compute.



\section{The TDA and RPA methods.}
\label{AP:TDA-and-RPA}
The RPA method can be seen as a direct extension of the
TDA method where the collective $ph$-operator is written
 (using the notation of Eq. (\ref{pair-states})) as
\beqa \label{TD-phonon}
\hat \Gamma^\dag_{n;\Gamma\mu}
=\sum_{{\bf a}, {\bf b}} X^n_{{\bf a}{\bf b};\Gamma}  
[{\bd b}^\dag_{\bf a} {\bd d}^\dag_{\bar {\bf b}} ]^\Gamma_\mu
~~~,
\eeqa
which restricts the approximation to the space of $1p-1h$ excitations,
relevant for low-lying states.
The $n$th TDA excited state is given by
$ |n;\Gamma\mu\rangle=\hat \Gamma^\dag_{n;\Gamma\mu} |\tilde 0 \rangle$, and the ground state
satisfies $\hat \Gamma^{n;\Gamma\mu} |\tilde 0 \rangle=0$.

The TDA equation of motion is
\beqa
\sum_{{\bf a}, {\bf b}}  
\langle \tilde 0 | 
\left[\left(  [{\bd b}^\dag_{{\bf a}'} {\bd d}^\dag_{\bar {\bf b}'} ]^{\Gamma'}_{\mu'}\right)^\dag, 
\left [   {\bd H}^{QCD}, [{\bd b}^\dag_{\bf a} {\bd d}^\dag_{\bar {\bf b}} ]^\Gamma_\mu \right ]  \right]
|\tilde 0 \rangle  
X^n_{{\bf a}{\bf b};\Gamma}
=E^{TDA}_{n;\Gamma'}  X^n_{{\bf a}'{\bf b}';\Gamma'}
\delta_{\Gamma'\Gamma}\delta_{\mu' \mu}~,
\eeqa
where the $1p-1h$ correlations are taken into account
 in the excited states, keeping the ground state $| \tilde 0 \rangle$ unchanged, and
\beqa
&&\left(  [{\bd b}^\dag_{{\bf a}} {\bd d}^\dag_{\bar {\bf b}}  ]^{\Gamma}_{\mu}\right)^\dag \nonumber\\
&&= \left( [{\bd b}^\dag_{\pi_{a} k_{a} j_{a}  Y_{a}T_{a} } \otimes 
{\bd d}^\dag_{\pi_{b} k_{b} j_{b}  \bar Y_{b},T_{b} }  ]^{J^P,(0,0),YT}_{M_J,~ 0, ~~T_z} 
\right)^\dag  \nonumber\\
&&= \sum_{m_a, c_a ,T_{az}}\sum_{m_b, c_b ,T_{bz}}
\langle j_a m_a,j_b -m_b \mid J M_J \rangle
\langle (10) c_a,(01)\bar c_b \mid (0,0)0 \rangle_1
\langle T_a T_{az},T_b -T_{bz} \mid T T_z \rangle\nonumber\\
&&\times
~{\bd d}^{\pi_b k_b j_b -m_b, \bar c_b \bar Y_b,T_b -T_{bz}}    
~{\bd b}^{\pi_a k_a j_a m_a, c_a, Y_a,T_a T_{az} } 
\nonumber\\
&&=
\langle  {\bf a} \mu_{\bf a},\bar {\bf b} \bar \mu_{\bf b} |  \Gamma \mu \rangle
{\bd d}^{\bar {\bf b} \bar \mu_{\bf b}}  {\bd b}^ {{\bf a}  \mu_{\bf a}}
~=~
[{\bd d}^{\bar {\bf b}}  {\bd b}^ {\bf a}  ]^\Gamma_\mu~~~,
\eeqa
where we have used the short-hand notation 
$\langle  {\bf a} \mu_{\bf a}, \bar {\bf b} \bar \mu_{\bf b} | \Gamma\mu \rangle$ 
for the product of the spin, color and isospin-flavor
Clebsch-Gordan coefficients. Notice that, whenever the coupling involves upper 
single-particle indices, we use the Clebsch-Gordan coefficients in the 
inverted order. This is done for convenience, in order to absorb phases.

The most straightforward generalization of the collective
$1p-1h$-operator is the RPA.
The phonon creation and annihilation operators are defined as
\beqa\label{RPA-phonon}
\hat \Gamma^\dag_{n;\Gamma\mu}
&=&\sum_{{\bf a}, {\bf b}} \left\{X^n_{{\bf a}{\bf b};\Gamma}  
[{\bd b}^\dag_{\bf a} {\bd d}^\dag_{\bar {\bf b}} ]^\Gamma_\mu
-Y^n_{{\bf a}{\bf b};\Gamma} 
(-1)^{\phi_{\Gamma\mu}}
[{\bd d}^{\bar {\bf b}}  {\bd b}^{ {\bf a}}  ]^{\bar \Gamma}_{\bar \mu}\right\}
\eeqa
and 
\beqa
\hat \Gamma^{n;\Gamma\mu} 
&=&\sum_{{\bf a},{\bf b}}
\left\{ \left(X^n_{{\bf a}{\bf b};\Gamma}\right)^*
[{\bd d}^{\bar {\bf b}}  {\bd b}^ {\bf a}]^\Gamma_\mu
-\left(Y^n_{{\bf a}{\bf b};\Gamma}\right)^* (-1)^{\phi_{\Gamma\mu}}
[{\bd b}^{\dag}_{\bf a} {\bd d}^{\dag}_ {\bar {\bf b} }  ]^{\bar \Gamma}_{\bar \mu} \right\}
~~~,
\eeqa
respectively.
The $n$th RPA state with quantum numbers $\Gamma\mu$ is constructed as
$| n; \Gamma\mu \rangle_{RPA} =\hat \Gamma^\dag_{n;\Gamma\mu}|RPA\rangle$. 
The RPA ground state  $|RPA\rangle$ is constructed in such a way that the condition
$\hat \Gamma^{n;\Gamma\mu}|RPA\rangle=0$ is fulfilled, {\it i.e.},
\beqa
&&| RPA \rangle \nonumber\\
&&= |\tilde 0\rangle 
+\sum_{n,\Gamma}\sum_{{\bf a}' {\bf b}' {\bf a} {\bf b}} 
Z^{n,\Gamma}_{{\bf a}' {\bf b}', {\bf a} {\bf b}} 
\left[   [{\bd b}^\dag_{{\bf a}'} {\bd d}^\dag_{\bar {\bf b}'} ]^\Gamma
\otimes
 [{\bd b}^\dag_{\bf a} {\bd d}^\dag_{\bar {\bf b}} ]^{\bar \Gamma} \right]^{0,(0,0),00}_{0,~0,~~0}
|\tilde 0\rangle \nonumber\\
&&= |\tilde 0\rangle 
+\sum_{n,\Gamma}\sum_{{\bf a}' {\bf b}' {\bf a} {\bf b}} 
Z^{n,\Gamma}_{{\bf a}' {\bf b}', {\bf a} {\bf b}} 
\sum_{\mu}
 \frac{(-1)^{\phi_{\Gamma\mu}}}{\sqrt{\mbox{dim}(\Gamma)}}
  [{\bd b}^\dag_{{\bf a}'} {\bd d}^\dag_{\bar {\bf b}'} ]^\Gamma_\mu
 [{\bd b}^\dag_{\bf a} {\bd d}^\dag_{\bar {\bf b}} ]^{\bar \Gamma}_{\bar\mu} 
|\tilde 0\rangle \nonumber\\
&&= |\tilde 0\rangle 
+\sum_{n,\Gamma}\sum_{{\bf a}' {\bf b}' {\bf a} {\bf b}} 
Z^{n,\Gamma}_{{\bf a}' {\bf b}', {\bf a} {\bf b}} 
\sum_{M_J T_z}
\frac{(-1)^{J-M_J} } {\sqrt{2J+1}}  \frac{(-1)^{T-T_z}}{\sqrt{2T+1}}
\nonumber\\
&&~~~~~~~~\times
[{\bd b}^\dag_{\pi_{a'} k_{a'} j_{a'}  Y_{a'}T_{a'} } \otimes 
{\bd d}^\dag_{\pi_{b'} k_{b'} j_{b'}  \bar Y_{b'},T_{b'} } ]^{J^P,(0,0),Y,T}_{M_J,~ 0, ~T_z}
[{\bd b}^\dag_{\pi_{a} k_{a} j_{a}  Y_{a}T_{a} } \otimes 
{\bd d}^\dag_{\pi_{b} k_{b} j_{b}  \bar Y_{b},T_{b} } ]^{J^P,(0,0),\bar Y,T}_{-M_J, 0, -T_z}
~|\tilde 0\rangle~,
\eeqa
with $ \sqrt{\mbox{dim}(\Gamma)}= \sqrt{2J+1} \sqrt{2T+1}$, 
$(-1)^{\phi_{\Gamma\mu}}=(-1)^{J-M_J} (-1)^{T-T_z} $
and  $Z^{n,\Gamma}_{{\bf a}' {\bf b}' , {\bf a} {\bf b}} =\sqrt{\mbox{dim}(\Gamma)}
Y^n_{{\bf a}' {\bf b}';\Gamma} \left(X^n_{{\bf a}{\bf b};\Gamma}\right)^{-1}$.

The equation of motion $( {\bd H}^{QCD}|n,\Gamma\mu \rangle = E^{RPA}_{n,\Gamma} |n,\Gamma\mu\rangle )$ 
in the RPA formalism is equivalent to the double commutator
\beqa\label{RPA-eq.motion}
\langle  RPA |  
\left[\hat \Gamma^{n';\Gamma\mu},\left[ {\bd H}^{QCD} ,\Gamma^\dag_{n;\Gamma\mu}\right]\right]
|RPA \rangle=E^{RPA}_{n;\Gamma} \delta_{n,n'}~,  
\eeqa
with eigenvalues $E^{RPA}_{n;\Gamma\mu} $. 
Therefore, from Eq. (\ref{RPA-eq.motion}) we get two sets of equations 
which in matrix form can be written
\beqa\label{RPA-eqs-compact}
\left(\begin{array}{c c} A & B  \\B^* & A^*   \\\end{array}\right)
\left(\begin{array}{c} X^n   \\ Y^n    \\ \end{array}\right) 
=E^{RPA}_{n}
\left(\begin{array}{c c} 1 & 0  \\0 & -1   \\\end{array}\right)
\left(\begin{array}{c} X^n   \\ Y^n    \\ \end{array}\right) ~,
\eeqa
with
\beqa\label{Ap:F-and-B}
A_{a'b';\Gamma'\mu';~a,b;\Gamma\mu}&=&
\langle \tilde 0 | \left[ [{\bd d}^{\bar {\bf b}'}  {\bd b}^ {{\bf a}'}]^{\Gamma'}_{\mu'},
\left[ {\bd H}^{QCD} ,  [{\bd b}^\dag_{\bf a} {\bd d}^\dag_{\bar {\bf b}} ]^\Gamma_\mu \right]  \right] 
| \tilde 0 \rangle ~, \nonumber\\
B_{a'b';\Gamma'\mu';~a,b;\Gamma\mu}&=&-
\langle \tilde 0 | \left[ [{\bd d}^{\bar {\bf b}'}  {\bd b}^ {{\bf a}'}]^{\Gamma'}_{\mu'},
\left[ {\bd H}^{QCD} ,(-1)^{\phi_{\Gamma\mu}}
[{\bd d}^{\bar {\bf b}}  {\bd b}^{ {\bf a}}  ]^{\Gamma}_{\bar \mu} \right] \right] 
| \tilde 0 \rangle 
\eeqa
being the forward and backward matrices of the RPA method,
respectively. They are given explicitly in Appendix \ref{appC}.

To calculate the commutators in Eq. (\ref{Ap:F-and-B}) we enforce the
 {\it quasi-boson} approximation
\beqa\label{quasi-boson}
\langle RPA | \left[ [{\bd d}^{\bar {\bf b}'}  {\bd b}^ {{\bf a}'}]^{\Gamma'}_{\mu'} , 
[{\bd b}^\dag_{\bf a} {\bd d}^\dag_{\bar {\bf b}} ]^\Gamma_\mu \right] | RPA \rangle 
\simeq
\langle \tilde 0| \left[ [{\bd d}^{\bar {\bf b}'}  {\bd b}^ {{\bf a}'}]^{\Gamma'}_{\mu'} , 
[{\bd b}^\dag_{\bf a} {\bd d}^\dag_{\bar {\bf b}} ]^\Gamma_\mu \right] | \tilde 0\rangle 
=\delta_{{\bf b} ' {\bf b}}\delta_{{\bf a}' {\bf a}} \delta_{\Gamma ' \Gamma}\delta_{\mu ' \mu}~.
\eeqa

\section{
The Forward and Backward matrices of the Effective QCD Hamiltonian.}
\label{appC}
The forward matrix elements are given by
\beqa
&&A_{a'b';\Gamma'\mu';~a,b;\Gamma\mu}\nonumber\\
&&=\delta_{\Gamma\Gamma'}\delta_{\mu\mu'}
\sum_{ k j \pi Y T }  \epsilon_{k j \pi Y T}\nonumber\\
&&\times ~\delta_{\pi_{a'}\pi_{a}}  \delta_{\pi_{b'}\pi_{b}}
\delta_{k_{a'}k_{a}}  \delta_{k_{b'}k_{b}}
\delta_{j_{a'}j_{a}}  \delta_{j_{b'}j_{b}}
\delta_{Y_{f_a'}Y_{f_a}}  \delta_{Y_{f_b'}Y_{f_b}}
\delta_{T_{f_a'}T_{f_a}}   \delta_{T_{f_b'}T_{f_b}}
\left(
\delta_{k k_{a}}\delta_{j j_{a}} \delta_{T_{f} T_{f_a}}\delta_{\pi \pi_{a}}
+
\delta_{k k_{b}}\delta_{j j_{b}} \delta_{T_{f} T_{f_a}}\delta_{\pi \pi_{b}}
\right)\nonumber\\
&&- ~\frac{1}{2} ~  \delta_{\Gamma\Gamma'}\delta_{\mu\mu'}
\sum_{\lambda_i k_i \pi_i j_i Y_i  T_i}\sum_{L}  
~\frac{\sqrt{8(2L+1)}}{9} V^{L}_{ \{\lambda_i \pi_i k_i j_i T_i  \} }\nonumber\\
&&
\times \bigg\{ ~~
(\delta_{\lambda_1,\frac{1}{2}} \delta_{\lambda_2,\frac{1}{2}}
\delta_{\lambda_3,\frac{1}{2}} \delta_{\lambda_4,\frac{1}{2}} )
(\delta_{\pi_1 \pi_{a'}}\delta_{\pi_b \pi_{b'}}\delta_{\pi_2  \pi_3}\delta_{\pi_4 \pi_a}) ~
( \delta_{k_1 k_{a'}}\delta_{k_b k_{b'}} \delta_{k_2 k_3}\delta_{k_4 k_a} ) ~
( \delta_{j_1 j_{a'}}\delta_{j_b j_{b'}} \delta_{j_2 j_3}\delta_{j_4 j_a} )~\nonumber\\
&&
\times( \delta_{T_1 T_{a'}}\delta_{T_b T_{b'}} \delta_{T_2  T_3}\delta_{T_4  T_a} )
(\delta_{T_1 T_2}\delta_{T_3T_4}) ~
( \delta_{Y_1 Y_{a'}}\delta_{Y_b Y_{b'}} \delta_{Y_2 Y_3}\delta_{Y_4  Y_a} )
(\delta_{Y_1 Y_2}\delta_{Y_3 T_4})  ~   (-1)^{L-j_2+j_a} \frac{ \delta_{j_a j_{a'}}}{2j_a+1} \nonumber\\
&&\nonumber\\
&&
+~~ (\delta_{\lambda_1,\frac{1}{2}} \delta_{\lambda_2,\frac{1}{2}}
\delta_{\lambda_3,-\frac{1}{2}} \delta_{\lambda_4,-\frac{1}{2}} )
(\delta_{\pi_1 \pi_{a'}}\delta_{\pi_4 \pi_{b'}}\delta_{\pi_2  \pi_a}\delta_{\pi_3 \pi_b}) ~
(\delta_{k_1 k_{a'}}\delta_{k_4 k_{b'}}\delta_{k_2  k_a}\delta_{k_3 k_b}) ~
(\delta_{j_1 j_{a'}}\delta_{j_4 j_{b'}}\delta_{j_2  j_a}\delta_{j_3 j_b})~\nonumber\\
&&
\times
(\delta_{T_1 T_{a'}}\delta_{T_4 T_{b'}}\delta_{T_2  T_a}\delta_{T_3 T_b}) 
(\delta_{T_1 T_2}\delta_{T_3T_4}) ~
(\delta_{Y_1 Y_{a'}}\delta_{Y_4 Y_{b'}}\delta_{Y_2 Y_a}\delta_{Y_3 Y_b}) 
(\delta_{Y_1 Y_2}\delta_{Y_3 T_4}) ~(-1)^{j_b-j_a+J} 
\left\{\begin{array}{lll} j_b & j_a & J \\ [0.01in] 
 j_{a'} & j_{b'} & L   \end{array}\right\} \nonumber\\
&&\nonumber\\
&&
+~~ (\delta_{\lambda_1,-\frac{1}{2}} \delta_{\lambda_2,-\frac{1}{2}}
\delta_{\lambda_3,\frac{1}{2}} \delta_{\lambda_4,\frac{1}{2}} )
(\delta_{\pi_3 \pi_{a'}}\delta_{\pi_2 \pi_{b'}}\delta_{\pi_4  \pi_a}\delta_{\pi_1 \pi_b}) ~
(\delta_{k_3 k_{a'}}\delta_{k_2 k_{b'}}\delta_{k_4  k_a}\delta_{k_1 k_b}) ~
(\delta_{j_3 j_{a'}}\delta_{j_2 j_{b'}}\delta_{j_4  j_a}\delta_{j_1 j_b})~\nonumber\\
&&
\times
(\delta_{T_3 T_{a'}}\delta_{T_2 T_{b'}}\delta_{T_4  T_a}\delta_{T_1 T_b}) 
(\delta_{T_1 T_2}\delta_{T_3T_4}) ~
(\delta_{Y_3 Y_{a'}}\delta_{Y_2 Y_{b'}}\delta_{Y_4 Y_a}\delta_{Y_1 Y_b}) 
(\delta_{Y_1 Y_2}\delta_{Y_3 T_4})
~(-1)^{j_b-j_a+J} 
\left\{\begin{array}{lll} j_b & j_a & J \\ [0.01in] 
 j_{a'} & j_{b'} & L   \end{array}\right\} \nonumber\\
&&\nonumber\\
&&
+~~ 
(\delta_{\lambda_1,-\frac{1}{2}} \delta_{\lambda_2,-\frac{1}{2}}
\delta_{\lambda_3,-\frac{1}{2}} \delta_{\lambda_4,-\frac{1}{2}} )
(\delta_{\pi_a \pi_{a'}}\delta_{\pi_2 \pi_{b'}}\delta_{\pi_1  \pi_4}\delta_{\pi_3 \pi_b}) ~
( \delta_{k_a k_{a'}}\delta_{k_2 k_{b'}} \delta_{k_1 k_4}\delta_{k_3 k_b} ) ~
( \delta_{j_a j_{a'}}\delta_{j_2 j_{b'}} \delta_{j_1 j_4}\delta_{j_3 j_b} )~\nonumber\\
&&
\times( \delta_{T_a T_{a'}}\delta_{T_2 T_{b'}} \delta_{T_1  T_4}\delta_{T_3  T_b} )
(\delta_{T_1 T_2}\delta_{T_3T_4}) ~
( \delta_{Y_a Y_{a'}}\delta_{Y_2 Y_{b'}} \delta_{Y_1 Y_4}\delta_{Y_3  Y_b} )
(\delta_{Y_1 Y_2}\delta_{Y_3 T_4})  ~   (-1)^{L-j_1+j_b} \frac{  \delta_{j_b j_{b'}}}{2j_b+1} \nonumber\\
&&\nonumber\\
&&
+ ~~ (\delta_{\lambda_1,-\frac{1}{2}} \delta_{\lambda_2,\frac{1}{2}}
\delta_{\lambda_3,\frac{1}{2}} \delta_{\lambda_4,-\frac{1}{2}} )
 (\delta_{\pi_3 \pi_{a'}}\delta_{\pi_b \pi_{b'}}\delta_{\pi_2  \pi_a}\delta_{\pi_1 \pi_4}) ~
(\delta_{k_3 k_{a'}}\delta_{k_b k_{b'}}\delta_{k_2  k_a}\delta_{k_1 k_4}) ~
(\delta_{j_3 j_{a'}}\delta_{j_b j_{b'}}\delta_{j_2  j_a}\delta_{j_1 j_4})~\nonumber\\
&&
(\delta_{T_3 T_{a'}}\delta_{T_b T_{b'}}\delta_{T_2  T_a}\delta_{T_1 T_4}) 
(\delta_{T_1 T_2}\delta_{T_3T_4}) ~
(\delta_{Y_3 Y_{a'}}\delta_{Y_b Y_{b'}}\delta_{Y_2 Y_a}\delta_{Y_1 Y_4}) 
(\delta_{Y_1 Y_2}\delta_{Y_3 T_4}) ~   (-1)^{L+j_1+j_a} \frac{ \delta_{j_a j_{a'}}}{2j_a+1} \nonumber\\
&&\nonumber\\
&&
+~~ (\delta_{\lambda_1,-\frac{1}{2}} \delta_{\lambda_2,\frac{1}{2}}
\delta_{\lambda_3,\frac{1}{2}} \delta_{\lambda_4,-\frac{1}{2}} )
 (\delta_{\pi_a \pi_{a'}}\delta_{\pi_4 \pi_{b'}}\delta_{\pi_2  \pi_3}\delta_{\pi_1 \pi_b}) ~
(\delta_{k_a k_{a'}}\delta_{k_4 k_{b'}}\delta_{k_2  k_3}\delta_{k_1 k_b}) ~
(\delta_{j_a j_{a'}}\delta_{j_4 j_{b'}}\delta_{j_2  j_3}\delta_{j_1 j_b})~\nonumber\\
&&
(\delta_{T_a T_{a'}}\delta_{T_4 T_{b'}}\delta_{T_2  T_3}\delta_{T_1 T_b}) 
(\delta_{T_1 T_2}\delta_{T_3T_4}) ~
(\delta_{Y_a Y_{a'}}\delta_{Y_4 Y_{b'}}\delta_{Y_2 Y_3}\delta_{Y_1 Y_b}) 
(\delta_{Y_1 Y_2}\delta_{Y_3 T_4}) ~   (-1)^{L+j_2+j_b} \frac{ \delta_{j_b j_{b'}}}{2j_b+1} 
~~\bigg\}~. \nonumber\\
\label{Ceq2}
\eeqa

The backward matrix elements are
\beqa
&&B_{a'b';\Gamma';~a,b;\Gamma}
~=~ -\frac{1}{2}~ \delta_{\Gamma\Gamma'}\delta_{\mu\mu'}
\sum_{\lambda_i k_i \pi_i j_i Y_i  T_i}\sum_{L}  
~\frac{\sqrt{8(2L+1)}}{9} V^{L}_{ \{\lambda_i \pi_i k_i j_i T_i  \} }\nonumber\\
&&
\times \{~~
(\delta_{\lambda_1,\frac{1}{2}} \delta_{\lambda_2,-\frac{1}{2}}
\delta_{\lambda_3,\frac{1}{2}} \delta_{\lambda_4,-\frac{1}{2}} )
(\delta_{\pi_3 \pi_{a'}}\delta_{\pi_2 \pi_{b'}}\delta_{\pi_1  \pi_a}\delta_{\pi_4 \pi_b}) ~
(\delta_{k_3 k_{a'}}\delta_{k_2 k_{b'}}\delta_{k_1  k_a}\delta_{k_4 k_b}) ~
(\delta_{j_3 j_{a'}}\delta_{j_2 j_{b'}}\delta_{j_1  j_a}\delta_{j_4 j_b})~\nonumber\\
&&
\times
(\delta_{T_3 T_{a'}}\delta_{T_2 T_{b'}}\delta_{T_1  T_a}\delta_{T_4 T_b}) 
(\delta_{T_1 T_2}\delta_{T_3T_4}) ~
(\delta_{Y_3 Y_{a'}}\delta_{Y_2 Y_{b'}}\delta_{Y_1 Y_a}\delta_{Y_4 Y_b}) 
(\delta_{Y_1 Y_2}\delta_{Y_3 T_4})
\left\{\begin{array}{lll} j_{b'} & j_{'a} & J \\ [0.01in] 
 j_b & j_a & L   \end{array}\right\}   \nonumber\\
&&\nonumber\\
&&
+
(\delta_{\lambda_1,\frac{1}{2}} \delta_{\lambda_2,-\frac{1}{2}}
\delta_{\lambda_3,\frac{1}{2}} \delta_{\lambda_4,-\frac{1}{2}} )
(\delta_{\pi_1 \pi_{a'}}\delta_{\pi_4 \pi_{b'}}\delta_{\pi_3  \pi_a}\delta_{\pi_2 \pi_b}) ~
(\delta_{k_1 k_{a'}}\delta_{k_4 k_{b'}}\delta_{k_3  k_a}\delta_{k_2 k_b}) ~
(\delta_{j_1 j_{a'}}\delta_{j_4 j_{b'}}\delta_{j_3  j_a}\delta_{j_2 j_b})~\nonumber\\
&&
\times
(\delta_{T_1 T_{a'}}\delta_{T_4 T_{b'}}\delta_{T_3  T_a}\delta_{T_2 T_b}) 
(\delta_{T_1 T_2}\delta_{T_3T_4}) ~
(\delta_{Y_1 Y_{a'}}\delta_{Y_4 Y_{b'}}\delta_{Y_3 Y_a}\delta_{Y_2 Y_b}) 
(\delta_{Y_1 Y_2}\delta_{Y_3 T_4})
\left\{\begin{array}{lll} j_{b'} & j_{'a} & J \\ [0.01in] 
 j_b & j_a & L   \end{array}\right\}  
~~\}~.\nonumber\\
\eeqa

\section{Renormalization procedure.}
\label{Ap-RN}
In this appendix we elaborate on the renormalization of the quark
masses $m_{u,d},~m_s$ and the interaction parameters $\alpha$ and $\beta$, 
{\it at least approximately}, when the cut-off, given by the maximal
number of oscillation quanta $N_{{\rm cut}} \geq N_0$, 
is changed. We have selected a low-energy renormalization point
denoted by $N_0=3$, therefore in this renormalization
procedure $N_{{\rm cut}}$ always increases.

After the diagonalization of the kinetic term (Eqs.\ (\ref{eq:prediag}) and (\ref{eq9})), the Hamiltonian has the
following structure
\beqa
{\bd H} & = & \sum_p \varepsilon_p {\bd n}_p - \alpha \frac{1}{{\bd r}}
+ \beta {\bd r}
~~~,
\label{Eeq1}
\eeqa
where ${\bd r}$ is a short-hand notation for the linear interaction,
$\frac{1}{\bd r}$ for the standard Coulomb interaction and $\varepsilon_p$
is the single-particle energy in the state $p$. The operator ${\bd n}_p$
is a short-hand notation for the sum of the quark and antiquark number operators
related to the state $p$.

The meson states we construct (approximately) fulfill
the Schr\"odinger equation ${\bd H}\mid \Psi_i\rangle$
= $E_i \mid \Psi_i\rangle$. We multiply the equation from the left with $\langle \Psi_i \mid$,
using the notation $\langle ... \rangle$ for $\langle \Psi_i\mid ... \mid
\Psi_i\rangle$, to arrive at
\beqa
\langle {\bd H} \rangle _{N_{\rm cut}} & = & 
\sum_p \varepsilon_p^{N_{\rm cut}} \langle {\bd n}_p \rangle _{N_{\rm cut}}
- \alpha _{N_{\rm cut}}\langle \frac{1}{{\bd r}} \rangle _{N_{\rm cut}}
+ \beta _{N_{\rm cut}} \langle {\bd r} \rangle _{N_{\rm cut}}
~~~,
\label{Eeq2}
\eeqa
where we have explicitly indicated the dependence of the expectation values
on the cutoff $N_{{\rm cut}}$.
The first contribution on the R.H.S. is the prediagonalized kinetic energy. 
Its eigenstates correspond to states of free motion and, if no boundary
condition is imposed, the eigenvalues have to be continuous. In this case the
harmonic oscillator basis is not good at all. However,  we consider
{\it confined} particles,
and fixing the energy of the first single-particle state
$\varepsilon_1$ to a definite value independent of $N_{\rm cut}$ 
(see Table \ref{RN-parameters}) amounts to defining a scale or
extension for the system, which is proportional to the harmonic
oscillator length $\frac{1}{\sqrt{B_0}}$.

Now assume that, for a given value of the cut-off $N_0$, the parameters of the Hamiltonian
are adjusted to reproduce some low lying state ({\it e.g.}, the pion state). 
The expectation value of the Hamiltonian is then
\beqa
\langle {\bd H} \rangle_{N_0} & = & 
\sum_p \varepsilon_p^{N_0} \langle {\bd n}_{p} \rangle_{N_0} 
- \alpha_{N_0} \langle \frac{1}{{\bd r}} \rangle_{N_0}
+\beta_{N_0} \langle {\bd r} \rangle_{N_0}
~~~.
\label{Eeq3}
\eeqa

Its relation to (\ref{Eeq2}) with a finite cut-off at $N_{{\rm cut}}$ is
\beqa
\langle {\bd H} \rangle_{N_{{\rm cut}}} & = & 
\langle {\bd H}\rangle_{N_0} + \langle {\bd H}\rangle_{N_{{\rm cut}}>N_0}
\label{Eeq3a}
\eeqa
where the last term refers to the contributions for $N_{{\rm  cut}}>N_0$. 
Equation (\ref{Eeq3a}) can be extended to any kind of operator, replacing in the equations above 
the Hamiltonian ${\bd H}$ with the operator ${\bd O}$, which can be ${\bd r}$
or any of those appearing in (\ref{Eeq1}).

The {\it renormalization condition} is that the (low-lying) energies, as the observables, do not
change when the cut-off is increased (or decreased). In order to achieve this,
we first discuss the renormalization of the masses and then the renormalization of the interaction parameters
$\alpha$ and $\beta$.

The mass parameters are contained implicitly in  the single-particle energies  $\varepsilon_p^{N_0}$. 
One contribution to $\varepsilon_p^{N_0}$ comes from the kinetic energy,
which is not subject to renormalization, and the other from 
the masses of the different types of quarks, which have to be renormalized.
The renormalization of the quark masses can be achieved by a 
multiplicative factor, {\it i.e.},
\beqa
m^{f}_{N_{{\rm cut}}} & \approx & 
m^{f}_{N_0} \sqrt{x^f_{{\rm N}_{{\rm cut}}}}
~~~,
\label{Eeq3cs}
\eeqa
where we have taken into account that the factor may depend on the
flavor of the quark, $q$ = u,d or s.

The purpose of Eq.\ (\ref{Eeq3cs}) is to guarantee that the effective single-particle
energies  ($m_q^{eff}=\varepsilon_{1,\frac{1}{2},\pm\frac{1}{2},\frac{1}{2}}$ and
$m_s^{eff}=\varepsilon_{1,\frac{1}{2},0,0}$)   remain unchanged as
functions of $N_{\rm cut}$, see Table \ref{RN-parameters}. 
Therefore, the renormalization of the interaction can be performed independently.

We shall now describe the renormalization of the interaction:\\
Let us take as an example the linear interaction operator; then (\ref{Eeq3a}), multiplied
by the interaction parameter, can be rewritten as
\beqa
\beta_{N_{{\rm cut}}}
\langle {\bd r} \rangle_{N_{{\rm cut}}} & = & \beta_{N_{{\rm cut}}} \left(
1 + \frac{\langle {\bd r}\rangle_{N_{{\rm cut}}>N_0}}{\langle {\bd r}\rangle_{N_0}}
\right)
\langle {\bd r}\rangle_{N_0}
\nonumber \\
& = & \beta_{N_0}\langle {\bd r}\rangle_{N_0}
~~~,
\label{Eeq3b}
\eeqa
hence the relation between $\beta_{N_0}$ and the $\beta_{N_{{\rm cut}}}$ is
\beqa
\beta_{N_{{\rm cut}}} & = & \beta_{N_0} / \sqrt{x_{N_{{\rm cut}}}}
\nonumber \\
\sqrt{x_{N_{{\rm cut}}}} & = &  \left(
1 + \frac{\langle {\bd r}\rangle_{N_{{\rm cut}}>N_0}}{\langle {\bd r}\rangle_{N_0}}
\right)
~~~.
\label{Eq3c}
\eeqa
This guarantees the same expectation value via the renormalization
of $\beta$ as a function of the cut-off.  Equation (\ref{Eq3c}) incorporates
the contributions from $N$ larger than $N_0$. 
The square root is introduced for convenience. An analogous argument applies to the $\frac{1}{{\bd r}}$ interaction, 
and assuming that the relation $<\frac{1}{{\bd r}}>$ $\approx$ $\frac{1}{<{\bd r}>}$ 
is fulfilled, we have $\alpha_{N_{{\rm cut}}} \approx \alpha_{N_{0}}\sqrt{x_{N_{{\rm cut}}}}$.

We have performed numerical diagonalizations, increasing $N$ from 3 to
13 in steps of 2 and  determining $x_{N_{{\rm cut}}}$ and
$x^f_{N_{{\rm cut}}}$ such that the first single-particle energy remains
approximately constant, see Table \ref{RN-parameters}. 
Changing the masses and interaction constants 
according to the above equations results in an approximately invariant
RPA spectrum. Even though  we cannot determine exact expressions for the renormalization functions
because of the highly non-linear structure of the equations, 
a rough estimate of the factors $x_{N_{{\rm cut}}}$ and $x^f_{N_{{\rm    cut}}}$  is given by
\beqa\label{RN-ansatz}
x_{N_{{\rm cut}}} & \approx & 1 + t_1 \left(N_{{\rm cut}}-N_0\right)^{t_2}
\nonumber \\
x^f_{N_{{\rm cut}}} & \approx & 1 + t_1^f \left(N_{{\rm cut}}-N_0\right)^{t_2^f}
~~~,
\label{Eeq6}
\eeqa
where $t_1,t_2, t_1^f, t_2^f$ are  constants listed in Table \ref{RN-fits}, 
and $x_{N_{{\rm cut}}}$ and $x^f_{N_{{\rm cut}}}$ have to satisfy the
boundary conditions $x_{N_0}=1$ and $x^f_{N_0}=1$, respectively.

\end{document}